# In-plane electronic anisotropy of underdoped "122" Fe-arsenide superconductors revealed by measurements of detwinned single crystals

I. R. Fisher[1,2], L. Degiorgi[3] and Z. X. Shen[1,2]

(June 8th 2011)

1. Geballe Laboratory for Advanced Materials and Department of Applied Physics, Stanford University, CA 94305

2. Stanford Institute for Materials and Energy Sciences, SLAC National Accelerator Laboratory, 2575 Sand Hill Road, Menlo Park, California 94025, USA

3. Laboratorium für Festkörperphysik, ETH - Zürich, CH-8093 Zürich, Switzerland

**Abstract**

The parent phases of the Fe-arsenide superconductors harbor an antiferromagnetic ground state. Significantly, the Néel transition is either preceded or accompanied by a structural transition that breaks the four fold symmetry of the high-temperature lattice. Borrowing language from the field of soft condensed matter physics, this broken discrete rotational symmetry is widely referred to as an Ising nematic phase transition. Understanding the origin of this effect is a key component of a complete theoretical description of the occurrence of superconductivity in this family of compounds, motivating both theoretical and experimental investigation of the nematic transition and the associated in-plane anisotropy. Here we review recent experimental progress in determining the intrinsic in-plane electronic anisotropy as revealed by resistivity, reflectivity and ARPES measurements of detwinned single crystals of underdoped Fe arsenide superconductors in the "122" family of compounds.

## 1. Introduction

The various closely related families of Fe-pnictide and chalcogenide superconductors are well known, and we shan't belabor an unnecessary introduction. Several recent reviews [Paglione & Greene 2010, Mazin 2010, Johnston 2010], and early seminal papers describe the occurrence of superconductivity in materials which have commonly become known as "1111" (for example LaFeAsO [Kamihara *et al* 2008]), "122" (for example $BaFe_2As_2$ [Rotter *et al* 2008a,b]), "111" (LiFeAs [Tapp *et al* 2008] and NaFeAs [Parker *et al* 2009]) and "11" (for example Fe(Se,Te) [Hsu *et al* 2008]). With the apparent exception of LaFePO and LiFeAs, all of these families comprise a stoichiometric parent compound that exhibits an antiferromagnetic ground state. And in all cases, the antiferromagnetism is either accompanied or preceded in temperature by a structural transition that breaks a discrete rotational symmetry of the high-temperature crystal lattice. Borrowing language from the field of soft condensed matter physics, this broken rotational symmetry (from $C_4$ to $C_2$) is widely referred to as an Ising nematic phase transition [Fradkin *et al* 2010]. It is the purpose of this short review to draw together results of recent resistivity, reflectivity and ARPES measurements that reveal the associated in-plane electronic anisotropy that develops as the materials are cooled through this structural transition. Our discussion is limited solely to the "122" family of compounds, for which the majority of the measurements have thus far been performed. It will clearly be important to extend these measurements to other families in the near future in order to establish generic features. A necessary prerequisite for obtaining such results has been the development of techniques to detwin crystals *in situ*, which we will briefly review. Several groups have contributed to the emerging story, and we will try to describe each set of results accurately, acknowledging that we are best acquainted with measurements from our own collaborations.

Before going any further, we should emphasize that the origin and significance of the structural transition has been widely discussed from the very earliest days after the basic properties of the "1111" family of Fe-superconductors were determined. From a theoretical perspective, it was quickly appreciated that spin fluctuations might play a key role in driving the nematic transition [Fang *et al* 2008, Xu *et al* 2008, Ma *et al* 2008, Mazin *et al* 2009]. Within a local-moment Heisenberg model of a square lattice with nearest and next-nearest neighbor superexchange $J_1$ and $J_2$ in the regime $J_2 \geq 0.5 J_1$, both quantum and thermal fluctuations contribute a biquadratic term to the effective spin Hamiltonian which breaks the perfect frustration of the square lattice, and introduces an Ising nematic order parameter [Chandra *et al* 1990, Fang *et al* 2008, Xu *et al* 2008]. The applicability of such a local moment description has of course been questioned [Johannes & Mazin 2009], and it has also been suggested that if such a model is used then biquadratic terms must be included in the effective spin Hamiltonian irrespective of the effects of frustration, and that these terms are close in magnitude to the bilinear contribution [Yaresko *et al* 2009]. Within a more itinerant framework, a Pomerancuk type of instability might lead to better nesting of electron and hole pockets [Zhai *et al* 2009], while interaction between elliptical electron pockets also naturally leads to the presence of a biquadratic term in the effective Hamiltonian, presaging nematic order [Eremin & Chubukov 2010, Fernandes *et al* 2011]. Alternatively, the orbital degeneracy associated with partial filling of $d_{xz}$ and $d_{yz}$ states has inspired pictures in which the orbital degree of freedom plays an equally important role [Kruger *et al* 2009, Lv *et al* 2009, Chen *et al* 2009, Yanagi *et al* 2010, Laad *et al* 2010, Valenzuela *et al* 2010, Bascones *et al* 2010]. At this stage it is unclear which of these pictures provides the best description of the actual material, motivating detailed quantitative measurements of the in-plane electronic anisotropy that develops in the nematic state.

Measurement of the in-plane electronic anisotropy is of course hampered by twin formation (Figure 1). On cooling through the critical temperature associated with the structural transition, $T_s$, the materials tend to form dense structural twins, corresponding to alternation of the orthorhombic *a* and *b* axes [Tanatar *et al* 2009]. For probes which cannot distinguish the *a* and *b* orthorhombic directions, this effect obscures any in-plane anisotropy measured on length scales greater than the average twin dimension, which can be as small as a few microns. Even so, experiments performed using twinned crystals have yielded several key results which reveal a large in-plane anisotropy in the antiferromagnetic (AFM) state. In terms of the magnetic properties, early inelastic neutron scattering experiments for $CaFe_2As_2$ revealed an anisotropic spin-wave dispersion that can be best fit by a $J_{1a}$- $J_{1b}$ – $J_2$ model [Zhao *et al* 2009], where $J_{1a}$ and $J_{1b}$, the nearest neighbor exchange constants along the orthorhombic *a* and *b* axes, differ both in magnitude and sign. Similar measurements of $Ba(Fe_{1-x}Co_x)_2As_2$ have revealed anisotropic magnetic excitations even for the optimally doped composition [Lester *et al* 2010, Li *et al* 2010], and more recent measurements for $BaFe_2As_2$ have shown the presence of anisotropic spin fluctuations for temperatures well above $T_N$ [Harriger *et al* 2011], although in both cases the overall $C_4$ symmetry of the crystal lattice is not broken because the anisotropic spin wave dispersion is centered at (0,π) and (π,0) in the 1Fe unit cell notation. In terms of the electronic structure, analysis of quasiparticle interference observed via Scanning Tunnelling Microscopy (STM) measurements also indicates a large anisotropy in the electronic dispersion in the AFM state [Chuang *et al* 2010]. In addition, a polarization-dependent laser ARPES study on twinned $BaFe_2As_2$ in combination with first-principle calculations suggested a two-fold electronic structure in the SDW state [Shimojima *et al* 2010], and another ARPES study of crystals of $CaFe_2As_2$ for which the incident beam size was comparable to the size of the structural twins, revealed evidence for a large electronic anisotropy below the Néel temperature $T_N$ [Wang *et al* 2010]. These various observations have motivated several groups to find ways to detwin crystals in order to study the in-plane electronic anisotropy by other techniques, and for temperatures spanning $T_s$.

Before describing the methods used to detwin single crystals, and the resulting in-plane anisotropy that measurement of these detwinned samples reveals, we briefly comment on the separation of the structural and magnetic transitions. For the parent "122" compounds, initial measurements indicated that the tetragonal-to-orthorhombic structural transition occurred at the same temperature as the Néel transition [Goldman *et al* 2008, Huang *et al* 2008, Zhao *et al* 2008]. More recently, high resolution resonant x-ray diffraction measurements for $BaFe_2As_2$ indicate that for this compound $T_s$ occurs 0.75 K above $T_N$ [M. G. Kim *et al* 2011]. Substitution on the Fe site, for example by Co which has been widely studied, results in a suppression of both $T_s$ and $T_N$, but significantly $T_N$ is suppressed more rapidly, resulting in a progressive separation of the two transitions. The temperature difference monotonically increases with increasing concentration of the substituent, at least until the top of the superconducting dome [Ni *et al* 2008a, Chu *et al* 2009, Lester *et al* 2009, Ni *et al* 2010, Nandi *et al* 2010]. The origin of the splitting of $T_N$ and $T_s$ with chemical substitution is at present not clear, but consideration of the effect of crystal-quality on the splitting of the transitions in CeFeAsO [Jesche *et al* 2010] implies that this effect might, at least in part, be associated with the strong in-plane disorder introduced by partial substitution on the Fe site.

## 2. Methods to detwin single crystals *in-situ*

For all of the known families of Fe-arsenide and chalcogenide superconductors, the structural transition occurs below room temperature, and it is therefore necessary to detwin single crystals *in-situ*. Two distinct methods have been employed thus far; application of uniaxial stress (either compressive [Chu *et al* 2010b, Ying *et al* 2010, Yi *et al* 2011, Dusza *et al* 2011] or tensile [Tanatar *et al* 2010, Kim *et al* 2011]) and application of an in-plane magnetic field [Chu *et al* 2010a, Xiao *et al* (2010)]. The former method is superior, at least for the materials considered to date, resulting in almost complete detwinning. The latter method results in only a modest change in the relative twin domain populations for $Ba(Fe_{1-x}Co_x)_2As_2$, but has several other advantages that warrant a brief description.

### 2.1 Uniaxial stress

$BaFe_2As_2$ and closely related compounds form natural crystal facets along (100) and equivalent planes, referenced to the tetragonal structure. The orthorhombic *a* and *b* axes are oriented at 45 degrees to the tetragonal *a* axes (Figure 1), so in order to apply uniaxial stress along the orthorhombic *a*/*b* directions, crystals must be cut in to rectilinear bars with the tetragonal *a*-axis oriented at 45 degrees to the edges of the sample. Single crystal x-ray diffraction can be used to confirm the orientation of the crystal axes with respect to the cut edges, with typical errors of 2 degrees or less. Application of uniaxial stress at 45 degrees to the tetragonal *a* axis favors one twin orientation over the other upon cooling through the structural transition; if the uniaxial stress is compressive, the shorter *b* axis is favored in the direction of the applied stress, whereas the longer *a* axis is favored for the case of tensile stress. Experimental manifestations of detwinning devices from various groups based on this simple concept are shown in Figure 2. The extent to which the crystals have been detwinned can be monitored *in-situ* by either x-ray diffraction (Figure 3) or optical imaging using polarized light (Figure 4). For the case of the cantilever beam device, it is possible to obtain an order of magnitude estimate of the stress required to fully detwin the crystals based on the deflection of the cantilever. Typical values are in the range 5-10 MPa, comparable to pressures used to mechanically detwin $La_{2-x}Sr_xCuO_4$ [Lavrov *et al* 2001].

It is important to appreciate that uniaxial stress also has an effect for temperatures *above $T_s$*. Specifically, although under ambient conditions the lattice has a four-fold symmetry above $T_s$, application of uniaxial stress fundamentally breaks this symmetry. Under such conditions, a finite in-plane anisotropy is anticipated in all physical properties, where the anisotropy now refers to directions parallel and perpendicular to the applied stress. Indeed, any nematic order parameter that we define (for example $(\rho_b-\rho_a)/(\rho_b+\rho_a)$, where $\rho_a$ and $\rho_b$ refer to the resistivity along the *a* and *b* axes, or any other quantity that represents the difference in the properties of the material in the *a* and *b* directions) is finite for all temperatures above $T_s$ for a crystal held under uniaxial stress. Hence, there is no longer a nematic phase transition, and the associated divergent behavior of thermodynamic quantities is consequently rounded. The situation is exactly analogous to the case of cooling a ferromagnet in a magnetic field, for which the applied field results in a finite magnetization (order parameter) for all temperatures, and for which the divergent susceptibility at the Curie temperature is broadened. This effect can be appreciated by considering the temperature derivative of the resistivity as a function of applied stress, shown in Figure 5 for the specific case of $Ba(Fe_{1-x}Co_x)_2As_2$ with $x = 0.045$. Under ambient conditions, two sharp features are observed in the resistivity derivative at 58 and 68 K that match the heat capacity [Ni *et al* 2008a, Chu *et al* 2009], and which correspond respectively to the magnetic and structural transitions determined by x-ray and neutron scattering [Lester *et al* 2009]. Increasing the applied stress (in this case by adjusting the screw located part way along the cantilever, shown in Figure 2(a)) results in a progressive broadening of the feature at $T_s$, and a finite change in the resistivity for temperatures *above $T_s$*. In contrast, for temperatures below $T_s$, once the pressure exceeds a threshold value, there is little additional change in the resistivity, indicating that the sample is fully detwinned and that there is little pressure dependence to the resistivity of a single domain. Finally, from a symmetry perspective, the Néel transition is unaffected by the application of uniaxial pressure (after all, the Néel transition occurs in the orthorhombic state for $T_N < T_s$), and empirically the value of $T_N$ is not affected by the modest pressures that are exerted on the crystals in order to detwin them (Figure 5), at least for the materials considered so far.

## 2.2 In-plane magnetic fields

In 2002, Y. Ando and coworkers demonstrated that an in-plane magnetic field can influence the relative twin population of lightly doped $La_{2-x}Sr_xCuO_4$ [Lavrov *et al* 2002], which also suffers an orthorhombic transition, but in this case at ~ 450 K. The effect is due to a large in-plane anisotropy of the susceptibility tensor, which persists well above $T_N$ [Lavrov *et al* 2001]. Motivated by this example, similar experiments were attempted for $Ba(Fe_{1-x}Co_x)_2As_2$, revealing qualitatively similar effects below $T_N$ [Chu *et al* 2010a]. In this case, the susceptibility anisotropy has not been directly measured, but on quite general grounds it can be anticipated that the collinear antiferromagnetic structure that is adopted by the "122" Fe arsenides will result in a susceptibility that is larger for fields oriented along the *b*-direction (perpendicular to the orientation of the spins) than along the *a*-direction. The resulting difference in the energy of the two twin orientations in the presence of an in-plane field oriented along the *a/b* direction can be enough to move twin boundaries, as revealed by optical imaging (Figure 6). The change in the relative twin population is, however, only modest, of order 5 – 15% for typical laboratory scale fields. Moreover, the effect only appears to work below $T_N$ implying a much smaller susceptibility anisotropy above $T_N$ than is found for $La_{2-x}Sr_xCuO_4$.

Although the relative change in the twin population caused by the application of an in-plane magnetic field is small, the large in-plane anisotropy of the resistivity tensor of $Ba(Fe_{1-x}Co_x)_2As_2$ (described in detail below) results in a substantial magnetoresistance for temperatures below $T_N$ if the field is applied in the *ab*-plane (Figure 7) [Chu *et al* 2010a]. Rotation of the field within the *ab*-plane reveals a distinct hysteresis, also observed in field sweeps, associated with the twin boundary motion. This effect is thermally assisted and displays relaxation effects with a time constant that depends on temperature [Chu *et al* 2010a].

Typical laboratory-scale magnetic fields are unable to fully detwin crystals of $Ba(Fe_{1-x}Co_x)_2As_2$. Nevertheless, the effect has the distinct advantage of being continuously tunable in a straightforward manner, permitting experiments which explicitly probe the behavior of the twin domains. For example, rotation of the applied field in the *ab*-plane results in a hysteretic magnetoresistance (Figure 8(a)). By resolving the projection of the field separately on to the *x* and *y* axes, it is possible to construct a hysteresis loop for the twin boundary motion (Figure 8(b)). This is entirely analogous to the case of domain wall motion in a ferromagnet, but in this case associated with an Ising nematic system, for which $B_x^2-B_y^2$ plays the equivalent role that *B* plays in a ferromagnet.

In contrast to $Ba(Fe_{1-x}Co_x)_2As_2$, Xiao et al (2010) have demonstrated via a series of neutron scattering measurements that $EuFe_2As_2$ can be completely detwinned by application of modest in-plane magnetic fields. In this case, the field couples to the large Eu magnetic moment, inducing a metamagnetic transition to a saturated paramagnetic state with a critical field a little below 1 T. Apparently the large difference in Zeeman energy of the two twin orientations in the saturated paramagnetic state is sufficient to drive a complete detwinning of $EuFe_2As_2$, but the effect is presumably restricted to temperatures below the Néel temperature of the Eu sub-lattice, which is 20 K. Furthermore, the effect is not hysteretic, such that the twin structure reappears when the field is swept back to zero and the Eu ions resume their antiferromagmetic structure. Nevertheless, this fortuitous effect presents an opportunity to study the in-plane electronic anisotropy of $EuFe_2As_2$ at low temperatures via magnetotransport measurements, though to date we are unaware of any such experiments.

### 3. In-plane resistivity anisotropy

As described above, the relative population of twin domains of $Ba(Fe_{1-x}Co_x)_2As_2$ can be influenced by an in-plane magnetic field. The resulting magnetoresistance is not insignificant, even for relatively small changes in the twin population, from which it is possible to infer a rather large in-plane resistivity anisotropy with $\rho_b > \rho_a$ [Chu *et al* 2010a]. However, the magnetoresistance is rapidly suppressed above $T_N$ (Figure 7), and mechanically detwinned crystals provide a much better method to explore the intrinsic in-plane anisotropy. At the time of writing, the in-plane resistivity anisotropy has been investigated for mechanically detwinned crystals of the parent "122" compounds $AFe_2As_2$ (A = Ca, Sr, Ba) [Tanatar *et al* 2010, Blomberg *et al* 2010]; for the electron-doped systems $Ba(Fe_{1-x}Co_x)_2As_2$ [Chu *et al* 2010b, Liang *et al* 2011], $Ba(Fe_{1-x}Ni_x)_2As_2$, $Ba(Fe_{1-x}Cu_x)_2As_2$, [Kuo *et al* 2011] & $Eu(Fe_{1-x}Co_x)_2As_2$ [Ying *et al* 2010]; and for the hole-doped system $Ba_{1-x}K_xFe_2As_2$ [Ying *et al* 2010]. These measurements reveal several intriguing and unanticipated results which we discuss in greater detail below.

For all of the parent "122" compounds $AFe_2As_2$ (A=Ca, Sr, Ba, Eu) it is found that $\rho_b > \rho_a$, as anticipated by the magnetoresistance measurements of $Ba(Fe_{1-x}Co_x)_2As_2$. The reader is reminded that the *b*-axis is the shorter of the two in-plane lattice parameters, as well as being the direction in which the moments align ferromagnetically, so at first glance this observation seems somewhat counterintuitive. However, the magnitude of the anisotropy for $BaFe_2As_2$ is only modest, with maximal values of $\rho_b/\rho_a \sim$ 1.2 for temperatures close to $T_N$ [Chu *et al* 2010b]. The anisotropy is smaller still for $SrFe_2As_2$ and $CaFe_2As_2$ [Blomberg *et al* 2010]. The temperature dependence of the anisotropy through $T_N$ for the strained crystals is somewhat different for the four cases (Figure 9, Blomberg *et al* 2010), depending on the degree to which the coupled structural/magnetic transition is first or second order, to which we return shortly. Furthermore, absolute values of the anisotropy can vary depending on annealing conditions, implying a sensitivity to disorder [Liang *et al* 2011].

The in-plane resistivity anisotropy was also measured for $Ba(Fe_{1-x}Co_x)_2As_2$ for underdoped, optimally doped and overdoped compositions (Figure 10, Chu *et al* 2010b). As found for the undoped parent compounds, $\rho_b > \rho_a$ for all underdoped compositions. The temperature dependence of $\rho_a$ remains metallic to the lowest temperatures, whereas $\rho_b$ rapidly develops a steep upturn with decreasing temperature as the dopant concentration is increased from zero. Partial suppression of the superconductivity by an applied magnetic field oriented along the *c*-axis reveals that $\rho_b$ continues to rise with decreasing temperature for these underdoped compositions (Figure 11). Consideration of the derivative of the resistivity reveals that the in-plane anisotropy develops most rapidly for temperatures close to $T_s$ (see Figure 4(b) of Chu *et al* 2010b), implying that it is closely associated with the nematic transition. The most striking aspects of the data are both the large magnitude of the anisotropy, and the non-monotonic doping dependence, both of which are best appreciated by plotting the resistivity anisotropy $\rho_b/\rho_a$ as a function of temperature and composition (Figure 12). Both results are in stark contrast to the structural orthorhombicity, characterized by $(a-b)/(a+b)$, which is small (0.4% for $x$ = 0 at 7 K) and monotonically decreases with increasing Co concentration [Prozorov *et al* 2009]. In contrast, the in-plane resistivity anisotropy rises to a maximum value $\rho_b/\rho_a \sim 2$ for $x \sim 3.5\%$ [Chu *et al* 2010b]. Perhaps coincidentally, the large resistivity anisotropy appears to develop for compositions close to the onset of the superconducting dome. This is also close to the composition at which ARPES measurements indicate the vanishing of a hole-like pocket of reconstructed Fermi surface [Liu *et al* 2010]. We return to the possible significance of this Lifshitz transition in Section 6.

Inspection of Figures 10 and 12 reveals that the difference in $\rho_b$ and $\rho_a$ begins at a temperature *well above $T_s$*. Depending on composition, and on the magnitude of the applied stress, anisotropy in the resistivity can be observed up to ~ 2 $T_s$. As mentioned previously, in this temperature regime, the value of the anisotropy, and also the temperature to which the anisotropy persists, depends sensitively on the magnitude of the applied stress (Figures 5 & 24) [Chu *et al* 2010b, Liang *et al* 2011]. There is no indication in either the resistivity or its derivatives of an additional phase transition marking the onset of this behavior in either stressed or unstressed crystals. Furthermore, diffraction measurements reveal that for unstressed conditions the material is fundamentally tetragonal within the available resolution [Goldman *et al* 2008, Huang *et al* 2008, Zhao *et al* 2008, Kim *et al* 2011]. Hence, the simplest explanation of the observed anisotropy above $T_s$ for the stressed crystals is that this effect is induced by the uniaxial stress, rather than this being related to the presence of static nematic order. Within linear response theory, such an effect can be described in terms of a large *nematic* susceptibility that couples to uniaxial lattice strain. Courtesy of the fluctuation-dissipation theorem, this implies a wide temperature range above $T_s$ where nematic fluctuations are appreciable, with possible implications for the superconducting pairing mechanism. Resonant UltraSound measurements of the elastic moduli of $Ba(Fe_{1-x}Co_x)_2As_2$ also reveal a softening of the sheer modulus, consistent with the presence of a substantial nematic susceptibility [Fernandes *et al* 2010].

It is instructive to compare the strain-induced anisotropy for temperatures above $T_s$ of $Ba(Fe_{1-x}Co_x)_2As_2$, for which the structural transition is argued to be second order [Wilson *et al* 2009 & 2010, M. G. Kim *et al* 2011], with the case of $CaFe_2As_2$, for which the coupled magnetic/structural transition is strongly first order [Ni *et al* 2008b]. Specifically, uniaxial stress does not seem to cause an appreciable change in the resistivity above $T_s$ for $CaFe_2As_2$ [Tanatar *et al* 2010], consistent with the reduced effect of fluctuations anticipated for a first order phase transition. Comparison with $SrFe_2As_2$, for which the 1[st] order transition is somewhat weaker [Yan *et al* 2008] reveals that the induced resistivity anisotropy above $T_s$ is proportionately larger (Figure 9, Blomberg *et al* 2010).

It is of course important to ask to what extent the compositional dependence of the in-plane resistivity anisotropy observed for $Ba(Fe_{1-x}Co_x)_2As_2$ is generic. Measurements by Ying *et al* (2010) reveal a similarly large in-plane resistivity anisotropy for $Eu(Fe_{1-x}Co_x)_2As_2$. And our own more recent measurements of $Ba(Fe_{1-x}Ni_x)_2As_2$ (Figure 12) and $Ba(Fe_{1-x}Cu_x)_2As_2$ reveal a similar non-monotonic doping dependence to the anisotropy [Kuo *et al* 2011]. Thus, for electron-doped "122" systems which involve substitution on the Fe site, the appearance of a large resistivity anisotropy for some intermediate range of compositions seems to be generic, at least for the variants studied thus far. In stark contrast, measurements of detwinned single crystals of the hole-doped system $Ba_{1-x}K_xFe_2As_2$ (Figure 14, Ying *et al* 2010) indicate a vanishingly small in-plane anisotropy for $x = 0.1$ and $0.18$. It is not yet clear how the anisotropy of the parent compound $BaFe_2As_2$ evolves to give this essentially isotropic response for the K-doped samples, but taken at face value the result provides evidence that the effects of hole-doping and/or substitution off the FeAs plane might be rather different to electron-doping and/or substitution on the FeAs plane . We return to this point later.

In addition to the zero-field resistivity measurements described above, improvements in crystal quality have enabled the observation of Shubnikov-de Hass oscillations in the magnetoresistance of the parent compound $BaFe_2As_2$ over a wider field range than was previously possible [Analytis *et al* 2009], enabling a clearer identification of the distinct frequencies and their various harmonics. In a beautiful set of measurements using detwinned single crystals of $BaFe_2As_2$, Terashima *et al* (2011) have been able to map the reconstructed Fermi surface, revealing the presence of small isotropic pockets, as well as somewhat larger, more anisotropic pockets. We return to the significance of these observations in relation to the resistivity anisotropy in Section 6.

## 4. Anisotropic charge dynamics

The observation of a large in-plane resistivity anisotropy, at least for the electron-doped "122" Fe arsenides, bears witness to the orthorhombicity of the material, but does not distinguish between anisotropy in the electronic structure and anisotropy in the scattering rate. To this end, reflectivity measurements of detwinned single crystals using polarized light can provide important insight to the effects of the magnetic and structural transitions on the anisotropic charge dynamics and the electronic band structure. Both the overall spectral weight distribution and also the scattering rate of itinerant charge carriers may be extracted from analysis of the metallic contribution to the excitation spectrum. Such measurements for detwinned single crystals of Ba(Fe$_{1-x}$Co$_x$)$_2$As$_2$ have established a direct link between the rather counterintuitive temperature dependence of the anisotropic *dc* transport properties described above and the related charge dynamics in the underdoped regime [Dusza *et al* 2011]. These observations also enable a clearer understanding of the average optical response previously obtained for twinned crystals of the same material [Lucarelli *et al* 2010].

Exploiting a similar concept as described in Section 2.1 (Fig. 2a), a detwinning device (Fig. 2c) that allows optical measurements under constant uniaxial pressure was developed [Dusza *et al* 2011]. The device consists of a mechanical clamp and an optical mask attached on top and in tight contact. The pressure-device was designed according to the following specific criteria: i) it leaves the (001) facet of the single domain samples exposed, enabling optical reflectivity ($R(\omega)$) measurements; ii) the optical mask guarantees data collection on surfaces of the same dimension for the sample (S) and the reference mirror (M) and therefore on equivalent flat spots; iii) the uniaxial stress is applied by tightening a screw and drawing the clamp against the side of the crystal, cut such that in the orthorhombic phase the *a/b* axes of the twinned samples would lie parallel to the stress direction; iv) the major axis of the tightening screw lies nearby and parallel to the surface of the sample so that the shear- and thermal-stress effects are minimized. The thermal expansion $\Delta L$ of the tightening screw, exerting the uniaxial pressure, can be estimated to be of the order of $\Delta L = \alpha L \Delta T = 20$ μm (for screw-length L=5 mm, typical metallic thermal expansion coefficient $\alpha=2\times10^{-5}$ K$^{-1}$, and thermal excursion $\Delta T= 200$ K). This corresponds to a relative variation of about 0.4%. By reasonably assuming $\Delta L/L = \Delta p/p$, the influence of the thermal expansion is then negligible.

Prior to inserting the sample holder into the cryostat the alignment conditions between M and S were verified by imaging on both spots a red laser point source. Within the cryostat a micrometer allows a very accurate positioning of S and M with respect to the light beam. $R(\omega)$ data were thus collected as a function of temperature on these detwinned, single domain samples in the far- and mid-infrared (FIR and MIR) spectral range between 30 and 6000 cm$^{-1}$ [Dressel and Gruner 2002]. Data were complemented with room temperature measurements from the near-infrared (NIR) up to the visible and ultra-violet spectral range (3200-4.8x10$^4$ cm$^{-1}$). Light in all spectrometers was polarized along the *a* and *b* axes of the detwinned samples. The polarizers chosen for each measured frequency range have an extinction ratio greater than 200, thus reducing leakages below our 1% error limit. The real part $\sigma_1(\omega)$ of the optical conductivity was obtained via the Kramers-Kronig transformation of $R(\omega)$ by applying suitable extrapolations at low and high frequencies. For the $\omega \rightarrow 0$ extrapolation, the Hagen-Rubens (HR) formula ($R(\omega)=1-2\sqrt{(\omega/\sigma_{dc})}$) was used, inserting $\sigma_{dc}$ values in fair agreement with Chu *et al* (2010b), while above the upper frequency limit $R(\omega) \sim \omega^{-s}$ (2<s< 4) [Dressel and Gruner 2002]. Prior to performing optical experiments as a function of the polarization of light, the electrodynamic response of the twinned (i.e., unstressed) samples was first checked with unpolarized light, consistently recovering the same spectra previously presented by Lucarelli *et al* (2010). Although the detwinning device does not permit a precise control of the applied pressure, the uniaxial stress was carefully increased enough to observe optical anisotropy, which was verified to disappear when the pressure was subsequently released. As a control measurement, optical reflectivity data were also collected for a Cu sample of the same size (surface area and thickness) as that of the pnictide crystals and in the same set up with the uni-axial pressure. As expected, there is a total absence of dichroism for the stressed Cu sample, implying a vanishingly small sensitivity of the electronic structure to the modest uniaxial stresses employed in this measurement. The two compositions of $Ba(Fe_{1-x}Co_x)_2As_2$ with $x = 0$ and $0.025$ displayed overall similar features in their optical response [Dusza *et al* 2011]. To avoid repetition, raw data are shown only for $x = 0$. The discussion of the resulting optical anisotropy with respect to the *dc* findings is then discussed for both compositions.

The real part $\sigma_1(\omega)$ of the optical conductivity of the parent compound at 10 and 150 K along both polarization directions is shown in Fig. 15. Consistent with previous data on twinned samples [Lucarelli *et al* 2010], $\sigma_1(\omega)$ is dominated by a strong absorption peaked at about 4300 cm$^{-1}$ and a pronounced shoulder at 1500 cm$^{-1}$ on its MIR frequency tail. Inspection of Figure 15 reveals a distinct optical anisotropy that extends up to energies that far exceed the energy scales set by the transition temperatures. A similar optical anisotropy was also detected for the parent compound at 5 K by an independent investigation [Nakajima *et al* 2011]. In addition, the optical anisotropy of the stressed crystals is found to persist well above $T_s$, similar to the observation of an induced *dc* resistivity anisotropy for stressed crystals (Section 3). The anisotropy in the optical response for the magnetic state can be anticipated by *ab-initio* calculations based on density-functional-theory (DFT) as well as dynamical mean-field theory (DMFT) [Yin *et al* 2011, Sanna *et al* 2011, Sugimoto *et al* 2011]. It was first shown that the optical anisotropy of the magnetic state, not present within the local spin density approximation, may result from DMFT-correlation [Yin *et al* 2011]. Alternatively, DFT-calculations of the optical conductivity within the full-potential linear augmented plane-wave method reproduce most of the observed experimental features, in particular an anisotropic magnetic peak located at about 0.2 eV, which was ascribed to antiferromagnetically ordered stripes [Sanna *et al* 2011].

The inset in Fig. 15 shows an expanded view of the FIR $\sigma_1(\omega)$ well above and below $T_N$ for both *a* and *b* axes. With decreasing temperature, an overall enhancement of $\sigma_1(\omega)$ is observed in the FIR along the *a*-axis and its depletion along the *b*-axis. This reinforces the scenario for the opening of a pseudogap in the excitation spectrum for E∥*b* at $T_N$, reminiscent of what has been observed in twinned samples [Lucarelli *et al* 2010]. As will be discussed in greater detail below, such a depletion removes spectral weight, which principally piles up in the MIR feature at 1500 cm$^{-1}$. These findings demonstrate that the magnetic transition at $T_N$ seems to partially gap the portion of the Fermi surface pertinent to the *b*-axis response, while enhancing the metallic nature of the charge dynamics for the *a*-axis response.

A detailed analysis of the excitation spectrum was performed within the same Drude-Lorentz approach, previously introduced for the twinned samples [Lucarelli *et al* 2010] and adapted to both polarization directions. Figure 16 emphasizes the most relevant fit-components, pertinent for the spectral range characterized by a temperature dependent anisotropic optical response. The metallic part of $\sigma_1(\omega)$ consists of a narrow (N) and broad (B) Drude term; the former is obviously tied to the zero frequency extrapolation of R($\omega$), while the latter largely dominates the optical response. The two Drude terms imply the existence of two electronic subsystems [Wu *et al* 2010] and phenomenologically mimic the multi-band scenario in the iron-pnictides. A broad h.o. is used to describe the MIR energy interval (Fig. 16) [Lucarelli *et al* 2010], which accounts for the so-called magnetic-peak in $\sigma_1(\omega)$. Three high frequency Lorentz harmonic oscillators (h.o.) finally shape the interband transitions with onset at $\omega \sim 2000$ cm$^{-1}$. Such an analysis allows the spectral weight (SW) distribution to be disentangled with respect to the various energy intervals among all fit components, adding up in order to reproduce the complete excitation spectrum. The spectral weight is here defined in units of cm$^{-2}$ as SW=$(120/\pi)\int\sigma_1^i(\omega)d\omega$, where $\int\sigma_1^i(\omega)d\omega$ corresponds to the area of the $i^{th}$ component (Drude term or Lorentz harmonic oscillator) shaping $\sigma_1(\omega)$ (Fig. 16). Physically, the quantity SW corresponds to the square of the plasma frequency for the Drude terms or to the oscillator strength for the Lorentz components.

Figure 17 displays for both directions the total spectral weight, obtained from the sum of all oscillator strengths contributing to $\sigma_1(\omega)$ (Fig. 16), and its redistribution in terms of spectral weight for the Drude terms, the MIR-band as well as the high frequency electronic interband transitions. The total SW is constant along both polarization directions, thus satisfying the well-known *f*-sum rule [Dressel and Gruner 2002]. For E//*a* (Fig. 17, upper panel), the spectral weight moves from energies around the peak at 4300 cm$^{-1}$ into the broad MIR shoulder at 1500 cm$^{-1}$ (Fig. 15) and also down to low energies into the metallic contribution upon cooling below $T_N$. For E//*b* (Fig. 17, lower panel) on the other hand, there is a reshuffling of weight, which is lost by the Drude terms and piles up into the MIR-band. A similar trend in the SW reshuffling is also observed for $x = 0.025$, although somewhat less pronounced than for the parent compound. In order to emphasize the temperature dependence of the total Drude weight, we display the comparison between the two polarization directions in Fig. 18 for both compositions. Across the structural/magnetic phase transitions there is an important enhancement/depletion of the Drude weight with decreasing temperature along the *a*- and *b*-axis respectively, indicative of a reconstruction of the Fermi surface in the orthorhombic antiferromagnetic phase. The behavior of the Drude weight above $T_s$ is also interesting. It saturates to a constant value and gets larger along the *b*-axis than along the *a*-axis for $x = 0$, while for $x = 0.025$ it just merges into a constant and equal value for both directions. Figure 19 shows the scattering rates (i.e., the width at half maximum of the Drude terms [Dressel and Gruner 2002], Fig. 16) of the itinerant charge carriers for both narrow ($\Gamma_N$) and broad ($\Gamma_B$) Drude terms. Data are only displayed here for $x = 0$, but these are representative for all compositions in the underdoped regime. $\Gamma_N$ and $\Gamma_B$ increase along the antiferromagnetic *a*-axis, while they decrease along the ferromagnetic *b*-axis for $T < T_N$, as expected. Indeed, the large scattering rate along the *a*-axis may arise because of scattering from spin-fluctuations with large momentum transfer (i.e., by incoherent spin waves) [Turner *et al* 2009, Chen *et al* 2010]. The temperature dependence of the scattering rates paired with the non-negligible changes at energy scales close to the Fermi level shapes the *dc* transport properties (see below).

It is interesting to compare the anisotropic optical response with the anisotropy ratio of the *dc* transport properties, defined here as $\Delta\rho/\rho = 2(\rho_b-\rho_a)/(\rho_b+\rho_a)$. From the Drude terms the *dc* limit of the conductivity ($\sigma_0^{opt}=(\omega_{pN})^2/(4\pi\Gamma_N)+(\omega_{pB})^2/(4\pi\Gamma_B)$) for both axes) can be estimated more precisely than simply extrapolating $\sigma_1(\omega)$ to zero frequency. The anisotropy ratio $\Delta\rho^{opt}/\rho$, reconstructed from the optical data (i.e., $\sigma_0^{opt}$), is compared to the equivalent quantity from the transport investigation in Fig. 20. The agreement in terms of $\Delta\rho/\rho$ between the optical and *dc* investigation is outstanding for $x = 0.025$ at all temperatures (Fig. 20, lower panel). In comparison, $\Delta\rho^{opt}/\rho$ for $x = 0$ is slightly larger than the *dc* transport anisotropy for $T<T_s$ (Fig. 20, upper panel). This difference might originate from a difference in the applied stress in the optical and *dc* transport measurements, or from differences in the scattering rate of samples used for the two measurements (see discussion in section 6). Significantly, analysis of the optical reflectivity for all compositions indicates that anisotropy in the Fermi surface parameters, such as the enhancement(depletion) of the total Drude spectral weight occurring along the *a*(*b*)-axis (Fig. 18), *outweighs* the anisotropy in the scattering rates (Fig. 19) that develops below $T_N$ in terms of the effect on the *dc* transport properties. This is an important result from the optical investigation, which indeed enables both pieces of information governing the behavior of the *dc* transport properties to be extracted.

In order to emphasize the significant polarization dependence in $\sigma_1(\omega)$ at high frequencies, it is instructive to consider the difference $\Delta\sigma_1(\omega)=\sigma_1(\omega,E//a)- \sigma_1(\omega,E//b)$ at all measured temperatures. $\Delta\sigma_1(\omega)$ represents an estimation of the dichroism and turns out to be very prominent in the MIR and NIR ranges (Fig. 15, as well as Fig. 3d in Dusza *et al* 2011). It is especially interesting to compare the temperature dependence of the *dc* ($\Delta\rho/\rho$) and optical ($\Delta\sigma_1(\omega)$) anisotropy in the temperature range $T > T_s$ for which a large induced anisotropy is observed in the resistivity for stressed crystals (Section 3). Two characteristic frequencies are selected to follow the temperature dependence, identifying the position of the peaks in $\sigma_1(\omega)$ (Fig. 15); namely, $\omega_1=1500$ (1320) cm$^{-1}$ and $\omega_2=4300$ (5740) cm$^{-1}$ for $x = 0$ (0.025). It is remarkable that the temperature dependence of $\Delta\sigma_1(\omega)$ at $\omega_1$ and $\omega_2$ follows the temperature dependence of $\Delta\rho/\rho$ in both compounds (Fig. 20). $\Delta\sigma_1(\omega_i)$ (i=1,2) saturates at constant values well above $T_s$ and then displays a variation for $T < 2T_s$. The rather pronounced optical anisotropy, extending up to temperature higher than $T_s$ for the stressed crystals, clearly implies an important induced anisotropy in the electronic structure [Yin *et al* 2011, Sanna *et al* 2011, Sugimoto *et al* 2011], which is also revealed by ARPES measurements (Section 5). Since the dichroism directly relates to a reshuffling of spectral weight in $\sigma_1(\omega)$ in the MIR-NIR range (Fig. 15), $\Delta\sigma_1(\omega)$ at $\omega_1$ is interrelated to that at $\omega_2$, so that the behavior of $\Delta\sigma_1(\omega)$ is monotonic as a function of temperature and opposite in sign between $\omega_1$ and $\omega_2$ (Fig. 20). As anticipated, $\Delta\sigma_1(\omega_i) = 0$ for $T >> T_s$ for $x = 0.025$ (Fig 20, lower panel). However, for $x = 0$, $\Delta\sigma_1(\omega_i)$ is found to be constant but apparently different from zero for $T >> T_s$ (Fig 20, upper panel). The origin of the finite (but constant) dichroism at high temperatures for this sample is at present unclear, and might reflect a systematic effect due to imperfect experimental conditions (e.g., too strong applied uniaxial pressure). Nevertheless, the overall temperature dependence seems to behave in a very similar manner for the two compositions. Significantly, the absolute variation of the dichroism across the transitions at selected frequencies is larger for $x = 0$ than for $x = 0.025$. This doping-dependence needs to be studied in a controlled pressure regime in order to exclude effects arising from different degrees of detwinning ($T < T_s$) and different magnitude of induced anisotropy ($T > T_s$). Even so, it is encouraging to find evidence for changes in the electronic structure that, contrary to the *dc* resistivity [Chu *et al* 2010b], appear to follow the same doping-dependence as the lattice orthorhombicity [Prozorov *et al* 2009]. This behavior is clearly revealed in ARPES measurements of stressed crystals of Ba(Fe$_{1-x}$Co$_x$)$_2$As$_2$, described below.

## 5. Electronic anisotropy determined via ARPES

Adaptation of mechanical detwinning devices to allow cleaving of strained crystals *in situ* has recently enabled ARPES measurements on detwinned samples of BaFe$_2$As$_2$ [Y K Kim *et al* 2011] and Ba(Fe$_{1-x}$Co$_x$)$_2$As$_2$ [Yi *et al* 2011]. Hence, the band structure and reconstructed Fermi surface (FS) can be determined without the need to consider the effects associated with the superposition arising from different twin orientations. Furthermore, measurements performed for different orientations of the incident light polarization provide information about the dominant orbital characters of the bands via the photoemission matrix elements. Finally, these measurements also reveal the effect of uniaxial stress on the band structure for temperatures above $T_s$.

The Fermi surfaces of twinned and detwinned BaFe$_2$As$_2$ crystals in the SDW state at 10 K are compared in Fig 21 [Yi *et al* 2011]. Measurements were made using 25 eV photon energy and the FS was determined using an integration window of 5 meV about $E_F$. The Brillouin zone (BZ) is labeled corresponding to the true crystallographic 2-Fe unit cell in which Γ-X is along the antiferromagnetic crystal axis and Γ-Y is along the ferromagnetic (FM) crystal axis. As can be clearly seen, in the twinned case (Fig. 21(a)), the nearly orthogonal domains mix signals from both Γ-X and Γ-Y directions, masking any intrinsic differences between these directions and leading to a very complex FS topology and band dispersion. However, once detwinned, one clearly observes that the reconstructed FS appears very different in the two orientations (Figures 21(b&c)). Further measurements performed for different photon polarizations enable a more complete mapping of the reconstructed FS, and show that the FS remains multi-orbital in the SDW state [Yi *et al* 2011].

The most anisotropic features on the FS are the bright spots along Γ-X and petals along Γ-Y. Corresponding band dispersions (not shown here; see Yi *et al* 2011) reveal that these features arise from anti-crossing between hole and electron bands. Moreover, the bands cross at different energies in the two directions: close to $E_F$ along Γ-X inducing tiny Fermi pockets observed as bright spots, and 30meV below $E_F$ along Γ-Y resulting in bigger electron pockets on the FS. The area of these small pockets, corresponding to approximately 0.2% and 1.4% of the paramagnetic BZ, are in reasonable agreement with frequencies reported from quantum oscillation measurements [Analytis *et al* 2009]. These features have been observed in earlier ARPES data on twinned BaFe$_2$As$_2$ [Yi et al 2009, Richard *et al* 2010, Liu *et al* 2010] and SrFe$_2$As$_2$ [Hsieh *et al* 2008]. However, measurements of detwinned crystals clearly reveal that the bright spots and petals reside along two orthogonal directions, manifesting the orthorhombic symmetry of the electronic structure in the SDW state. Similar results were obtained by Kim *et al* [2011].

The most anisotropic feature in the low temperature state is a pronounced hole-like dispersion near the X and Y points, which evolves from the degenerate hole-like dispersion centred at the same points in the paramagnetic state (Figure 22). Consideration of the polarization dependence of the matrix elements [Yi *et al* 2011] allows an identification of the orbital character of these bands, which turn out to be principally $d_{yz}$ along Γ-X and $d_{xz}$ along Γ-Y, consistent with orbital assignments by non-magnetic LDA in the paramagnetic state [Graser *et al* 2010].

For temperatures well above $T_s$, cuts along Γ-X and Γ-Y reveal an identical electronic structure, as anticipated for the tetragonal crystal symmetry (where X and Y refer here to directions perpendicular and parallel to the compressive uniaxial stress). As temperature is lowered towards $T_{SDW}$, the $d_{yz}$ band along Γ-X shifts up and crosses $E_F$ whereas the $d_{xz}$ band along Γ-Y shifts down in energy, resulting in an unequal occupation of the $d_{yz}$ and $d_{xz}$ orbitals. The energy splitting between the originally degenerate $d_{yz}$ and $d_{xz}$ bands reaches ~60 meV at 80 K for the parent compound BaFe$_2$As$_2$ [Yi *et al* 2011].

Since uniaxial stress is used to detwin the single crystals, it is natural to ask whether the applied stress affects the splitting of the $d_{yz}$ and $d_{xz}$ bands. Figure 22 (a) shows cuts along the Γ-X and Γ-Y directions taken at 60 K for a detwinned single crystal of Ba(Fe$_{1-x}$Co$_x$)$_2$As$_2$ with $x = 0.025$, for which $T_S = $ 99 K and $T_N = $ 94.5 K, together with similar data for an unstressed sample. Panels (b) and (c) show the energy distribution curves (EDCs) of the photoemission intensity and its second derivative for the momentum indicated by the vertical yellow line in panel (a). As can be seen, within the uncertainty of the measurement, the energy splitting of the two bands is identical in the stressed and unstressed crystals. It appears that the modest uniaxial stress that is employed is sufficient to detwin the samples, but does not otherwise significantly perturb the band structure for $T < T_s$. The magnitude of the splitting of the $d_{yz}$ and $d_{xz}$ bands was measured as a function of cobalt concentration, for both stressed and unstressed crystals at a temperature of 10 K (Figure 23). These measurements reveal that the band splitting is uniformly suppressed together with the lattice orthorhombicity, as anticipated and consistent with the doping dependence of the dichroism observed in the MIR $\Delta\sigma_1(\omega_1)$ and $\Delta\sigma_1(\omega_2)$ observed in reflectivity measurements for the same Co concentration (Figure 20(b) and Section 4).

Finally, we comment on the effect of uniaxial pressure for temperatures above $T_s$. For stressed samples a clear splitting of the $d_{yz}$ and $d_{xz}$ bands can be observed for temperatures well above $T_s$ (summarized in Figure 24 – details can be found in Yi *et al*, 2011). This is consistent with the observation of an induced resistivity anisotropy for stressed samples, and points to the presence of a large electronic nematic susceptibility whereby a small uniaxial pressure generates a large difference in the band structure in the directions parallel and perpendicular to the applied stress. In contrast, unstressed crystals reveal no splitting well above $T_s$, as anticipated for the tetragonal crystal symmetry. Closer inspection of the data reveal a possible broadening or splitting of the bands close to $T_s$. Unstressed crystals reveal no features in the resistivity or its derivative in this regime (Figure 24 panels (b) and (c)), and hence it is unlikely that there is another phase transition associated with the onset of static electronic nematic order above $T_s$. It seems more likely that any slight splitting of the bands observed in the regime close to $T_s$ reflects the effect of nematic fluctuations close to the phase transition, or perhaps the presence of a slight relaxation at the surface which affects $T_s$ relative to the bulk.

## 6. Discussion

At this stage it is perhaps helpful to briefly summarize the key results gleaned thus far from measurements of detwinned single crystals. *DC* transport measurements of the parent compounds $BaFe_2As_2$, $SrFe_2As_2$ and $CaFe_2As_2$ reveal a modest in-plane anisotropy for temperatures below $T_s$ [Chu *et al* 2010a,b, Tanatar *et al* 2010, Blomberg *et al* 2010, Liang *et al* 2011]. The resistivity is found to be larger along the shorter of the orthorhombic crystal axes, corresponding to the direction in which the moments align ferromagnetically. Substitution of Co, Ni or Cu suppresses the lattice orthorhombicity [Prozorov *et al* 2009], but in contrast the in-plane resistivity anisotropy is found to initially increase with the concentration of the substituent, before reverting to an isotropic in-plane conductivity once the structural transition is completely suppressed [Chu *et al* 2010b, Kuo *et al* 2011]. Perhaps coincidentally, the onset of the large in-plane anisotropy for the cases of Co and Ni substitution occurs rather abruptly at a composition close to the start of the superconducting dome. For temperatures above $T_s$, there is a remarkably large sensitivity to uniaxial pressure, leading to a large induced in-plane resistivity anisotropy that is not observed for overdoped compositions [Chu *et al* 2010b]. There is no evidence in thermodynamic or transport measurements for an additional phase transition above $T_s$ for unstressed crystals, implying that the induced anisotropy is the result of a large nematic susceptibility, rather than the presence of static nematic order. Despite similarities between Co, Ni and Cu substitution, preliminary measurements of the in-plane resistivity of $Ba_{1-x}K_xFe_2As_2$ indicate that the large anisotropy observed for the electron-doped cases may not be found for the hole doped system [Ying et al 2010]. Reflectivity measurements provide valuable insight to the origin of the transport anisotropy of $Ba(Fe_{1-x}Co_x)_2As_2$. Measurements of detwinned single crystals reveal large changes in the low-frequency Drude response on cooling through $T_s$ and $T_N$, with a pronounced dichroism [Dusza et al 2011]. For light polarized in the antiferromagnetic *a* direction, there is an increase in the scattering rate, but this is accompanied by a dramatic increase in the spectral weight that ultimately leads to a reduction in the *dc* resistivity, consistent with observations. For light polarized along the *b* direction, the dominant effect is a reduction in the spectral weight, consistent with the increase in the *dc* resistivity. Thus, it appears that the dominant effect on the *dc* transport in the regime below $T_N$ is associated with changes in the electronic structure, although the scattering rate is also clearly affected. It should also be appreciated that the dichroism extends to very high energies, clearly revealing that changes in the electronic structure are not confined to near the Fermi energy (in keeping with a strong-coupling description of the material) and is smaller for higher Co concentrations. In addition, ARPES measurements provide clear evidence of the associated changes in the electronic structure. Measurements of detwinned single crystals of $Ba(Fe_{1-x}Co_x)_2As_2$ reveal an increase (decrease) in energy of bands with dominant $d_{yz}$ ($d_{xz}$) character on cooling through $T_s$ [Yi et al 2011], leading to a difference in orbital occupancy. The splitting of the $d_{xz}$ and $d_{yz}$ bands is progressively diminished with Co substitution in $Ba(Fe_{1-x}Co_x)_2As_2$, reflecting the monotonic decrease in the lattice orthorhombicity $(a-b)/(a+b)$. For temperatures above $T_s$, the band-splitting can be induced up to rather high temperatures by uniaxial stress, consistent with the dichroism observed in the MIR by reflectivity experiments. Finally, quantum oscillations in the parent compound reveal that the reconstructed FS comprises several small pockets [Terashima et al 2011]. The smallest of these pockets is essentially isotropic in the *ab*-plane, but the other, larger, pockets are much more anisotropic.

For temperatures below $T_N$, the FS is reconstructed due to the antiferromagnetic wavevector. In this temperature regime, the reflectivity measurements described above clearly show that the surprising dc resistivity anisotropy ($\rho_b > \rho_a$) is principally determined by anisotropy in the low frequency Drude spectral weight (i.e. changes in the electronic structure close to the Fermi energy). The scattering rate is somewhat larger along the antiferromagnetic *a* direction, but this effect is outweighed by the substantial increase in spectral weight for E||a. Of equal or perhaps greater interest is the temperature regime *above* $T_N$ for which the FS is not reconstructed. For strained crystals of $Ba(Fe_{1-x}Co_x)_2As_2$, an induced in-plane resistivity anisotropy is evident to high temperatures. One would like to understand whether this also originates from the Fermi surface, perhaps due to the difference in the orbital occupancy revealed by ARPES measurements [Chen et al 2010, Lv and Phillips 2011], or from anisotropic scattering, perhaps associated with incipient spin fluctuations [Fernandes et al 2011]. Unfortunately the current reflectivity data do not permit a conclusive answer to this question. Data for $x = 0.025$ reveal a progressive suppression of the difference in the spectral weight for $E||b$ and $E||a$ as the temperature is increased above $T_N$. Similar to the behavior below $T_N$, the total Drude weight from the narrow and broad terms is larger for $E||a$, but the two polarizations rapidly merge into a polarization independent constant value approximately 30 K above $T_N$. However, for $x = 0$ the data appear to exhibit a large and nominally temperature-independent anisotropy in the spectral weight to very high temperatures which is larger for $E||b$ than for $E||a$. In both cases, $x = 0$ and $x = 0.025$, the scattering rate associated with the broad Drude term appears to remain anisotropic to temperatures considerably higher than the dc transport anisotropy. The origin of this effect, and also the different behavior of the total Drude weight for the two compositions, is currently unclear. These different behaviors in the parameters determining the dc properties motivate further experiments in the temperature regime above $T_N$ in order to definitively address the origin of the anisotropy above the magnetic phase transition.

One of the most striking aspects of the results described above, is the non-monotonic doping dependence of the in-plane resistivity anisotropy that is observed for Ba(Fe$_{1-x}$Co$_x$)$_2$As$_2$ and Ba(Fe$_{1-x}$Ni$_x$)$_2$As$_2$ (Figures 12 & 13). For a multiband system, the conductivity tensor is the sum of the contribution from each pocket of the Fermi surface. Conversely, if one particular pocket dominates the conductivity tensor, then the transport anisotropy will also be determined by the anisotropy of that particular pocket. In the case of BaFe$_2$As$_2$, the reconstructed FS comprises at least three distinct pockets [Terashima et al 2011]. The angle-dependence of at least one of the frequencies observed in quantum oscillation measurements seems to match that anticipated for the FS pockets that arise from the protected band crossing mentioned in Section 5 [Harrison and Sebastian 2009, Ran et al 2009]. Furthermore, the observation of a large linear magnetoresistance in this compound might also be due to the linear dispersion associated with such a Dirac point [Huynh et al 2010], implying that this pocket is near the quantum limit [Abrikosov 1998] and has a relatively high mobility. In a separate ARPES study, Richard et al (2010) found that for BaFe$_2$As$_2$ the Dirac point is very close to the Fermi energy, and the associated FS pocket has only a weak in-plane anisotropy, consistent with more recent Shubnikov de Hass measurements [Terashima et al 2011]. It is therefore possible that the relatively small in-plane anisotropy that is observed for the parent AFe$_2$As$_2$ compounds (A=Ca, Sr, Ba & Eu) reflects the presence of a high-mobility, isotropic pocket of reconstructed FS associated with this Dirac point. Significantly, for Ba(Fe$_{1-x}$Co$_x$)$_2$As$_2$ the onset of the large in-plane resistivity anisotropy occurs for a composition very close to that at which earlier ARPES measurements indicate the disappearance of this small pocket of reconstructed Fermi surface [Liu et al 2010]. Analysis of the doping-dependence of both the linear coefficient of the magnetoresistance, and also the Hall coefficient, reveals a progressive erosion of the contribution to the conductivity associated with this pocket over the same range of compositions [Kuo et al 2011]. Taken together, these observations suggest that the onset of the large in-plane resistivity anisotropy is correlated with the progressive suppression of the contribution to the conductivity arising from the Dirac pocket. That is, the anisotropy associated with the other pockets of reconstructed FS can only be appreciated once the contribution from the isotropic pocket is diminished. Within such a picture, the effect of annealing on the in-plane resistivity anisotropy in undoped BaFe$_2$As$_2$ (Liang et al 2011) might be understood in terms of changes in the relative contribution to the total conductivity from the different pockets of reconstructed FS, though further experiments are necessary to establish this. In light of this hypothesis, it is interesting to consider the case of Ba$_{1-x}$K$_x$Fe$_2$As$_2$, for which preliminary measurements seem to indicate an essentially isotropic resistivity [Ying et al 2010]. The difference between transition metal substitution and Ba substitution might reflect differences in the effect of electron vs hole doping on the reconstructed FS and/or differences in scattering from anisotropic spin-fluctuations [Fernandes et al 2011]. Equally, it is likely that substitution away from the FeAs plane will have a weaker effect on the elastic scattering rate, which in turn would mean that the relative contribution to the conductivity arising from the isotropic Dirac pockets would not be diminished as rapidly as for in-plane transition metal substitution, perhaps consistent with the more isotropic conductivity that is observed for Ba$_{1-x}$K$_x$Fe$_2$As$_2$ below $T_N$. Further measurements are clearly necessary to provide more detailed information about the evolution of the reconstructed FS for the hole doped system.

As mentioned in Section 1, the origin of the orthorhombic transition has been discussed from the closely related perspectives of spin fluctuations (i.e. spin-driven nematicity [Fang *et al* 2008, Xu *et al* 2008, Mazin & Johannes 2009, Fernandes *et al* 2010, Fradkin *et al* 2010]), and also in terms of a more direct electronic effect involving, for instance, the orbital degree of freedom [Kruger *et al* 2009,Lee *et al* 2009,Lv *et al* 2010,Chen *et al* 2010,Valenzuela *et al* 2010,Bascones *et al* 2010]. From the perspective of symmetry, the present measurements cannot distinguish between these related scenarios, and of course the observation of in-plane anisotropy is implicit in the symmetry of the orthorhombic phase. Nevertheless, the magnitude of the observed effects, including the band splitting observed in ARPES measurements, provides important quantitative tests for more detailed future theoretical descriptions of the nematic phase transition in this material. Similarly, the relative extent to which uniaxial stress and in-plane magnetic fields couple to nematic fluctuations for temperatures above $T_s$ could potentially distinguish between pictures that are based on spin-driven nematic order vs orbital order. Such a comparison requires quantitative measurements of the pressure dependence of the induced anisotropy, which at present are unavailable, but nevertheless are entirely feasible.

Arguably the most dramatic effect revealed by all of the measurements described above is the remarkable sensitivity to uniaxial pressure displayed by the electronic structure for temperatures *above $T_s$*. As mentioned above, quantitative measurements of the pressure dependence of the induced anisotropy (i.e. the nematic susceptibility) as a function of chemical substitution are not currently available, and would clearly be welcome. The current data for Co and Ni substituted $BaFe_2As_2$ appear to indicate that the induced anisotropy is larger (and observed to higher temperatures) as the doping level is increased from the parent compound. The effect persists up to optimal doping, but is rapidly suppressed on the overdoped side of the phase diagram. This apparently non-monotonic doping-dependence of the nematic susceptibility is perhaps suggestive of the presence of a quantum critical point in the phase diagram, but it remains to be seen whether nematic fluctuations play any role in the superconducting pairing mechanism.

Finally, we remind the reader that in this short review we have limited ourselves to a presentation of the current state of experimental efforts aimed at revealing the in-plane electronic anisotropy via measurements of detwinned crystals of "122" Fe arsenides. It is unclear at this stage which if any of the results found for this material will be generic for the Fe-based superconductors, motivating experiments in other closely related families of compounds.

**Acknowledgments**

This review is inspired by work from several groups, including those of the coauthors. We have endeavored to faithfully reference all of the relevant literature, and apologize in advance for any inadvertent omissions. From our own collaborations we are especially indebted to J.-H. Chu, H.-H. Kuo, J. G. Analytis, S. C. Riggs, M. Yi, D.-H. Lu, C.-C. Chen, A. P. Sorini, A. F. Kemper, B. Moritz, R. G. Moore, M. Hashimoto, W.-S. Lee, T. P. Devereaux, K. De Greve, P. L. McMahon, & Y. Yamamoto (Stanford); S.-K. Mo & Z. Hussain (LBNL); Z. Islam (ANL); and A. Dusza, A. Lucarelli, F. Pfuner, J. Johannsen (ETHZ). We are also especially grateful to S. Kivelson, D. J. Scalapino, P. Hirschfeld, S. Massidda, J. Schmalian, R. Fernandez and C. Homes for many helpful conversations. Work at Stanford is supported by the DOE, Office of Basic Energy Sciences, Division of Materials Science and Engineering. Part of the magnetotransport measurements were performed at the National High Magnetic Field

Laboratory, which is supported by NSF Cooperative Agreement No. DMR-0654118, by the State of Florida, and by the DOE. Work at ETHZ has been supported by the Swiss National Foundation for the Scientific Research within the NCCR MaNEP pool.

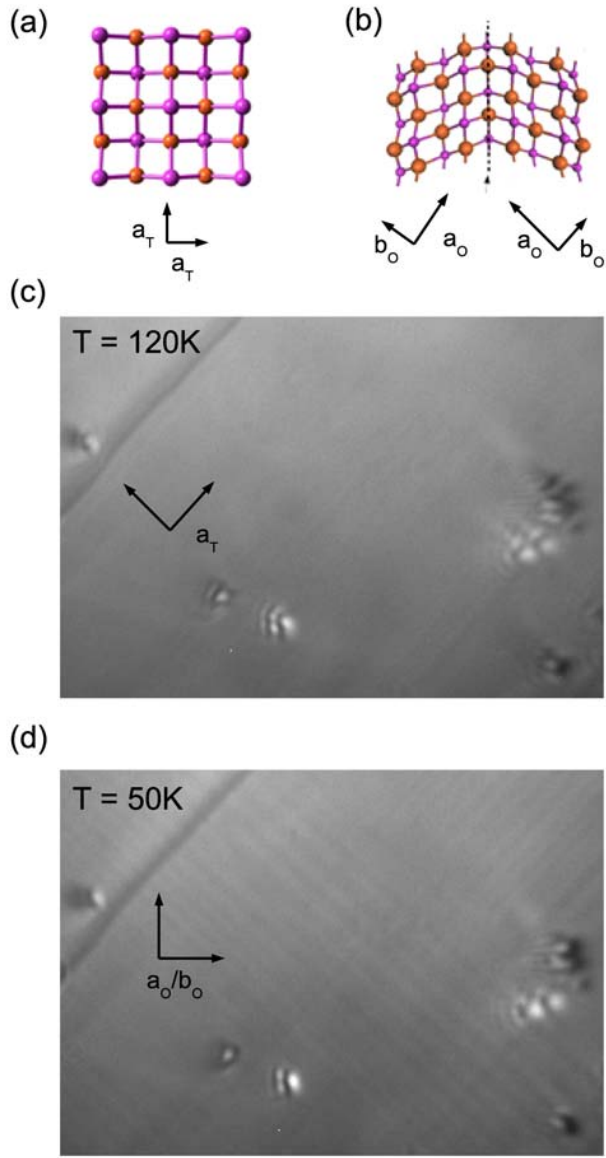

**Figure 1** Twin formation in underdoped Fe-arsenide superconductors. (a) Schematic diagram illustrating the Fe-As plane of $BaFe_2As_2$ in the tetragonal state, with the crystal axes labeled. Fe and As are shown as red and purple spheres respectively. (b) The structural phase transition corresponds to a stretching/contraction of the Fe-Fe distance along the orthorhombic $a$ and $b$ axes respectively. Twin boundaries separate regions for which the $a$ and $b$ crystal axes are oriented in opposite directions. The actual difference in the $a$ and $b$ lattice parameters is much less than illustrated in the diagram. (c,d) Optical images of a single crystal of $Ba(Fe_{1-x}Co_x)_2As_2$ with $x = 0.025$ at a temperature above and below the structural phase transition respectively. The images correspond to a region of the sample approximately 100 μm across, and were taken using almost crossed polarizers, such that the observed contrast reflects the different birefringence of the two twin orientations. Both images reveal the same surface morphology. Twin domains (regions of light or dark intensity) run diagonally from top left to bottom right, and from bottom left to top right in panel (d).

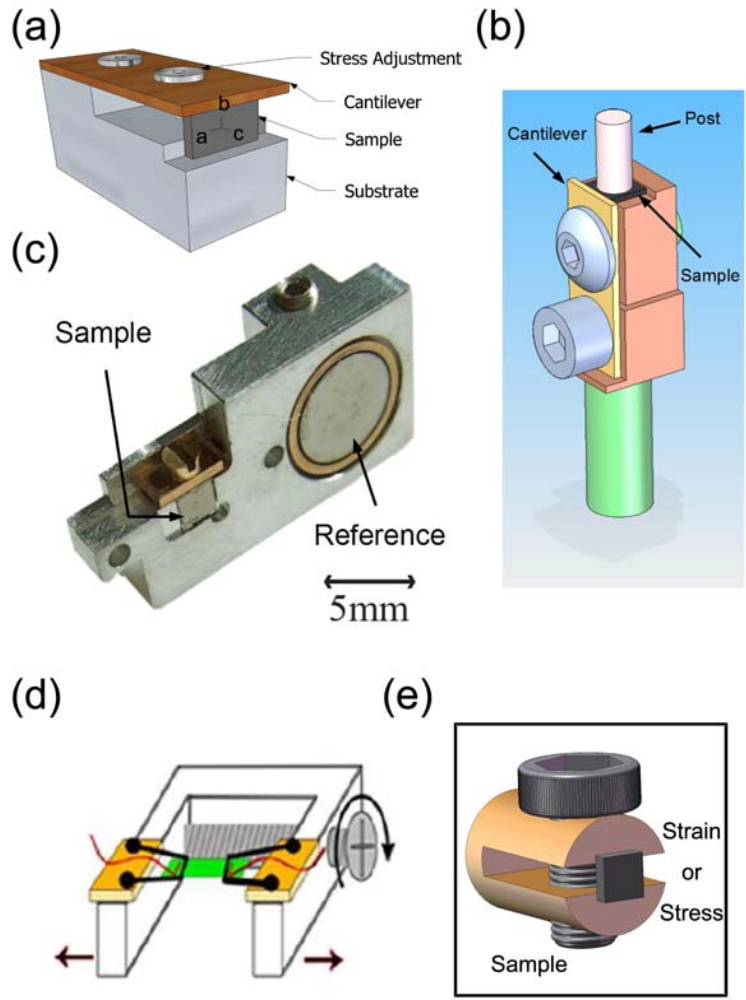

**Figure 2** Examples of experimental methods to mechanically detwin single crystals *in situ*. (a) A mechanical cantilever can be used to apply compressive stress. An order of magnitude estimate of the force applied to the crystal can be estimated from the curvature of the cantilever (From Chu *et al* (2010b). Reprinted with permission from AAAS.) Extensions of this basic idea include (b) attaching a post to the crystal to allow cleaving the strained crystal in situ for ARPES measurements, and (c) incorporating the detwinning device close to a reference metal for optical reflectivity measurements. (d) Equivalently, tensile stress provides an alternative means to mechanically detwin the crystals (reprinted with permission from Blomberg *et al* 2010, copyright (2010) by the American Physical Society). The elegant design shown in panel (e) enables both compressive and tensile stress to be applied, with the additional advantage that the sample is able to be cleaved *in situ* for surface sensitive measurements like ARPES (reprinted with permission from Y K Kim *et al* 2011, copyright (2011) by the American Physical Society). Apparently the applied stress is transmitted through the entire crystal to the exposed surface.

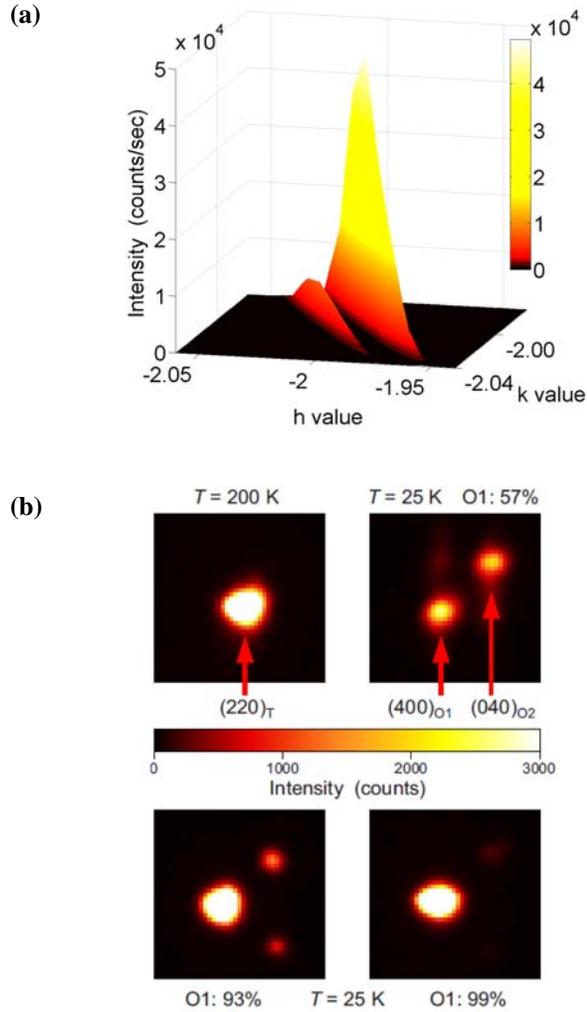

**Figure 3** Single crystal x-ray diffraction provides a direct measure of the relative population of the twin orientations below the structural transition. (a) The splitting of the $(-2\ -2\ 20)_T$ Bragg peak (referenced to the tetragonal lattice) at 40 K for a sample of $Ba(Fe_{1-x}Co_x)_2As_2$ with $x = 0.025$ under uniaxial pressure reveals a relative volume fraction of approximately 86%, with the shorter $b$ axis oriented along the direction of the compressive stress. The high photon energy and grazing angle of incidence ensure that almost the entire sample volume is probed (From Chu *et al* (2010b). Reprinted with permission from AAAS.) (b) Spatially resolved x-ray diffraction measurements for smaller $l$ indices can provide additional information about the degree of detwinning in individual parts of the strained crystal. Here, the authors show diffraction patterns for different regions of a single crystal held under tensile stress. The upper left panel shows the $(220)_T$ diffraction peak for $T > T_s$. The other three panels show the $(400)_{O1}$ and $(040)_{O2}$ diffraction peaks (referenced to the orthorhombic lattice) for different regions of the strained single crystal. The upper right hand panel is for a region close to the electrical contacts, for which the relative intensity of the two peaks are comparable in intensity. The two lower panels are for regions far from the electrical contacts, revealing 93 and 99% of the O1 twin orientation respectively (Reprinted with permission from Tanatar *et al* 2010, copyright (2010) by the American Physical Society).

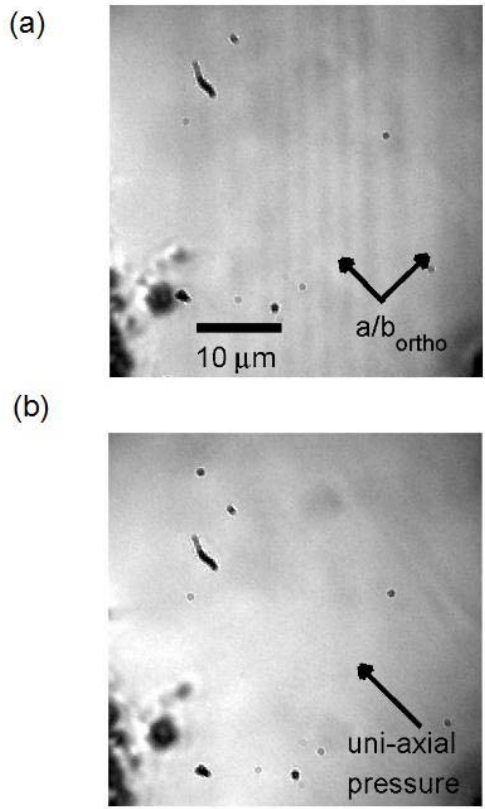

**Figure 4** Optical imaging using polarized light microcopy can also reveal the impact of uniaxial stress on the relative twin population. (a) An unstressed crystal of BaFe$_2$As$_2$ exhibits distinct twins that run vertically from top to bottom of the image. (b) Application of uniaxial stress via a cantilever detwinning device in the direction indicated leads to a single domain in the area imaged. (From Chu *et al* (2010b). Reprinted with permission from AAAS.)

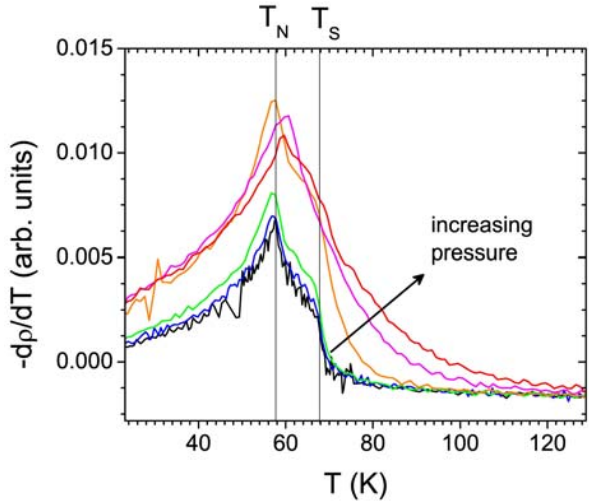

**Figure 5** Temperature-dependence of the resistivity derivative $d\rho/dT$ of a single crystal of Ba(Fe$_{1-x}$Co$_x$)$_2$As$_2$ with $x$ = 0.045 held in a cantilever detwinning device like that shown in Fig 2(a). Data are shown for a range of applied pressures, revealing the effect of uniaxial stress on the structural and magnetic phase transitions. Vertical lines indicate $T_N$ and $T_s$ determined in the absence of stress. Black curve is for ambient conditions, with no applied stress. Colored curves indicate the resistivity as the stress is progressively increased. Data are taken for current flowing in the direction of the applied stress. Below $T_s$, changes in the resistivity reflect detwinning of the crystal, which saturates beyond some value of the applied pressure. The Néel transition is unaffected by the applied stress, but the structural transition is rapidly broadened, affecting the resistivity well above $T_s$.

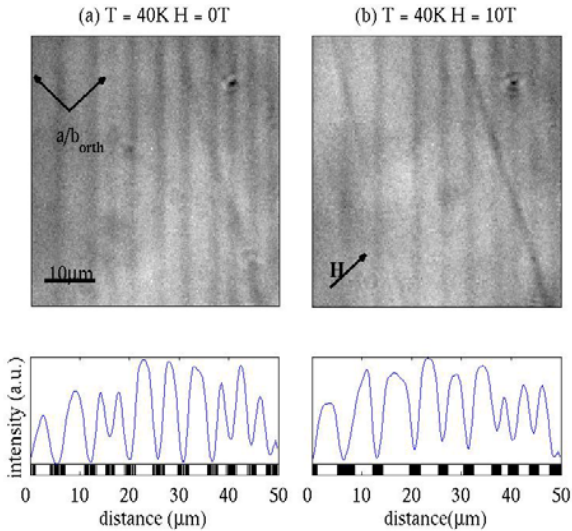

**Figure 6** Optical images of a Ba(Fe$_{1-x}$Co$_x$)$_2$As$_2$ sample for $x$=2.5%, revealing the partial detwinning effect of an in-plane magnetic field. The images were taken at $T$=40 K below both $T_s$ and $T_N$. The initial image (a) was taken in zero field, following a zero field cool from above $T_N$. The field was then swept to 10 T, at which field the second image (b) was taken. Horizontal intensity profiles, shown below each image, were calculated by integrating vertically over the image area after background subtraction and noise filtering. The field, which was oriented along the orthorhombic $a/b$ axes, has clearly moved the twin boundaries, favoring one set of twin domains (high intensity) over the other (low intensity). For the area shown, the relative fraction of domains with a high intensity changes from 54 ± 1 % to 61 ± 1 %. Reprinted with permission from Chu *et al* (2010a), copyright (2010) by the American Physical Society.

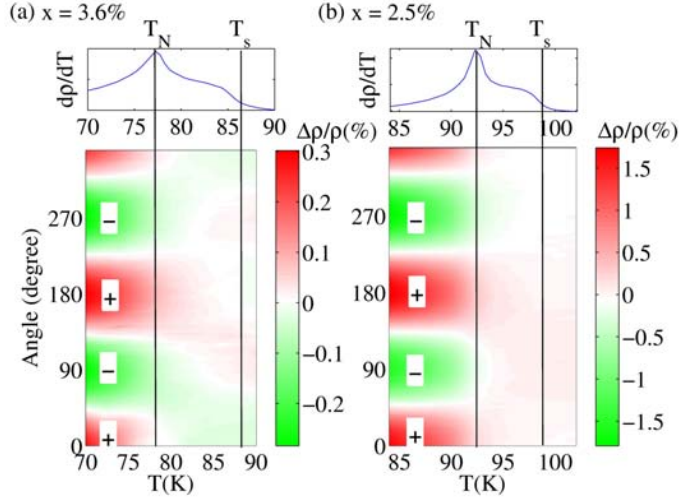

**Figure 7** In plane magnetoresistance of Ba(Fe$_{1-x}$Co$_x$)$_2$As$_2$ associated with twin boundary motion for temperatures close to $T_N$. Data were obtained by rotating a fixed magnetic field of 14 T in the *ab* plane. The crystals were cut in to rectilinear bars and the electrical contacts positioned such that the current is oriented along the *a/b* direction. Angles are measured with respect to the current direction. Vertical lines indicate $T_s$ and $T_N$, as determined for these samples by the derivative of the resistivity in zero field (upper panels), and by heat capacity and neutron scattering measurements for other samples. Reprinted with permission from Chu *et al* (2010a), copyright (2010) by the American Physical Society.

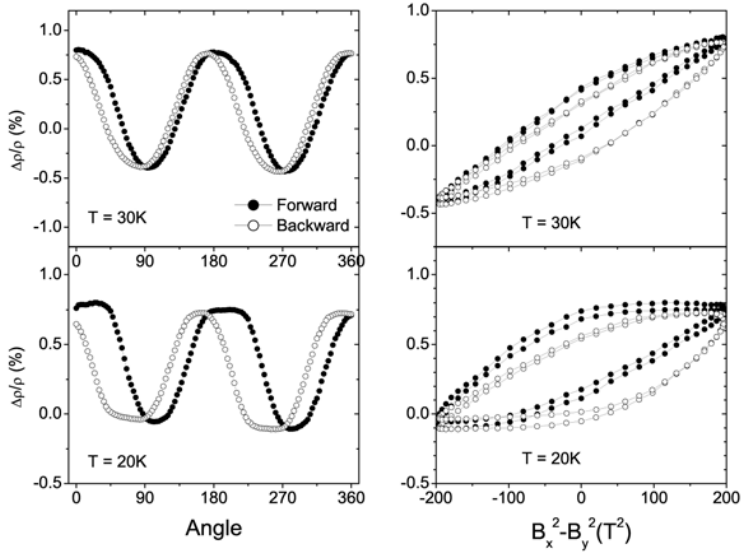

**Figure 8** Hysteresis of the in-plane magnetoresistance associated with twin boundary motion for a single crystal of Ba(Fe$_{1-x}$Co$_x$)$_2$As$_2$ with $x = 0.035$ well below $T_N$. As for figure 7, the current is oriented along the *a/b* direction, and angles are measured with respect to the current direction. Data were obtained by rotating a fixed field of 14 T in the *ab* plane, first in one direction (solid symbols), and then back again (open symbols). The data are plotted as a function of $B_x^2 - B_y^2$ in the right hand panel, revealing the hysteresis loops associated with twin boundary motion, analogous to the motion of ferromagnetic domains in a ferromagnet. (Figure courtesy J.-H. Chu)

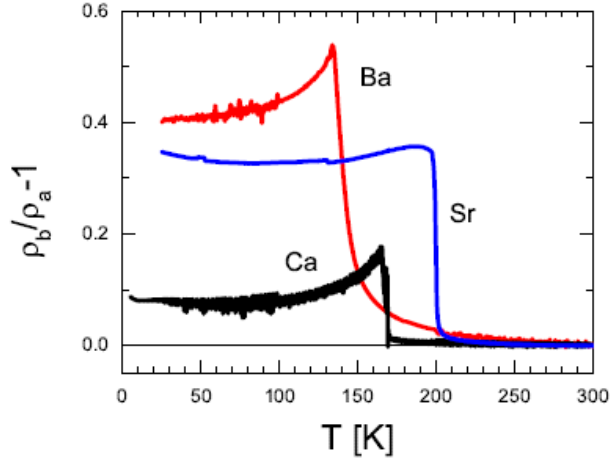

**Figure 9** Comparison of the in-plane anisotropy of CaFe$_2$As$_2$, SrFe$_2$As$_2$ and BaFe$_2$As$_2$. Note the larger induced anisotropy above $T_s$ for BaFe$_2$As$_2$ relative to the cases of SrFe$_2$As$_2$ and CaFe$_2$As$_2$. Although it is not known whether the same stress is applied for each measurement, nevertheless the larger induced anisotropy is consistent with the second order nature of the structural phase transition in BaFe$_2$As$_2$, relative to the more strongly first order transition observed for SrFe$_2$As$_2$ and CaFe$_2$As$_2$. Taken with permission from Blomberg *et al* (2010).

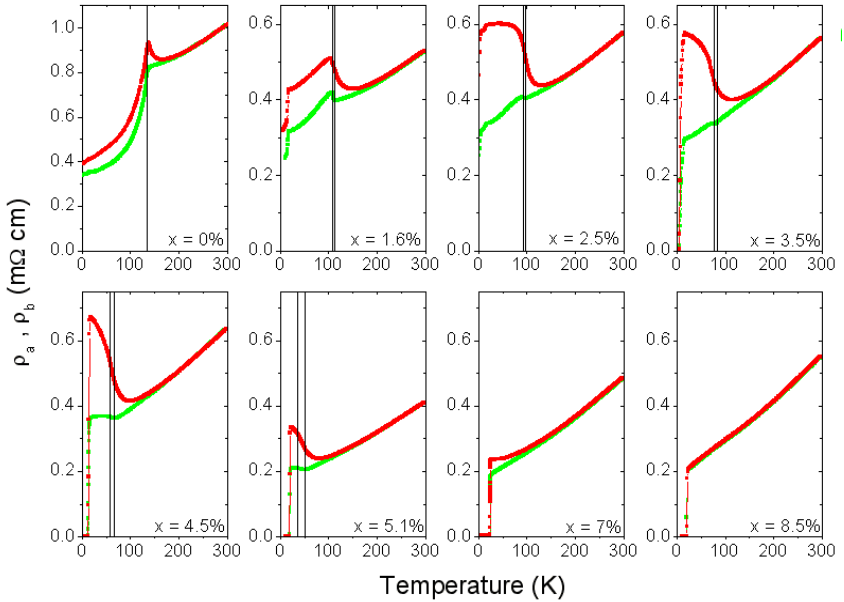

**Figure 10** In-plane resistivity anisotropy of Ba(Fe$_{1-x}$Co$_x$)$_2$As$_2$. Red lines show data for current parallel to applied uniaxial stress ($\rho_b$ configuration) and green lines show data for current perpendicular to stress ($\rho_a$ configuration). For each composition, measurements were made for the same crystal with the same set of contacts. Vertical lines show $T_N$ and $T_s$, determined in the absence of any stress. (From Chu *et al* (2010b). Reprinted with permission from AAAS.)

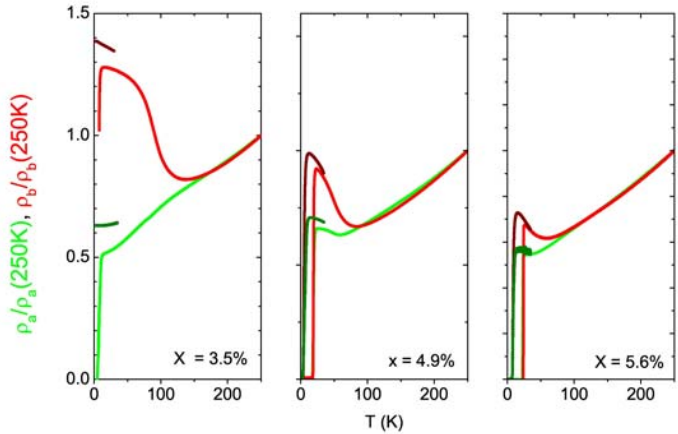

**Figure 11** Temperature dependence of the in-plane resistivity anisotropy of $Ba(Fe_{1-x}Co_x)_2As_2$ in the presence of a large magnetic field oriented along the $c$-axis to partially suppress the superconductivity. Red (green) lines show data for current parallel (perpendicular) to applied uniaxial stress corresponding to the $\rho_b$ ($\rho_a$) configuration in zero field. Dark red (dark green) data points show $\rho_b$ ($\rho_a$) in a field of 35 T. For all three compositions, $\rho_b$ continues to rise to the lowest temperatures, until cut off by the superconducting transition. (Figure courtesy J.-H. Chu)

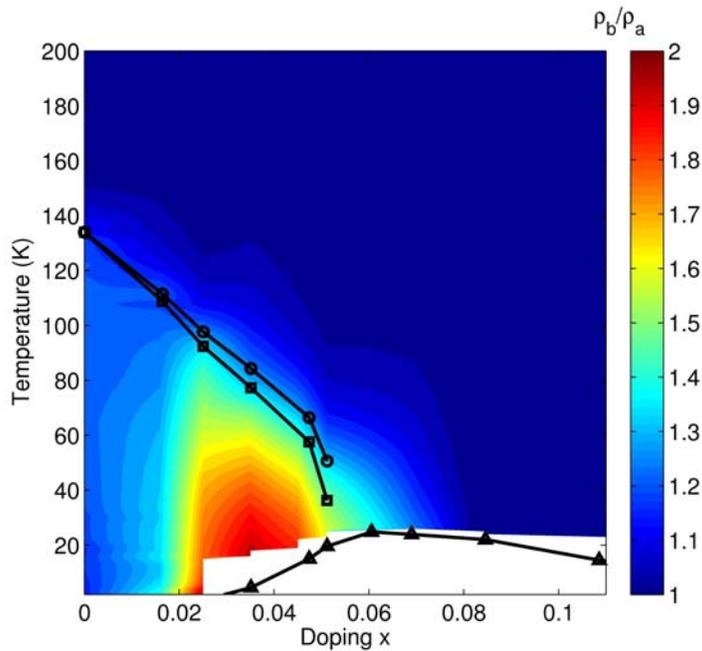

**Figure 12** Temperature and composition dependence of the in-plane resistivity anisotropy $\rho_b/\rho_a$ of $Ba(Fe_{1-x}Co_x)_2As_2$. Structural, magnetic and superconducting critical temperatures, determined in the absence of uniaxial stress, are shown as circles, squares, and triangles, respectively. (From Chu *et al* (2010b). Reprinted with permission from AAAS.)

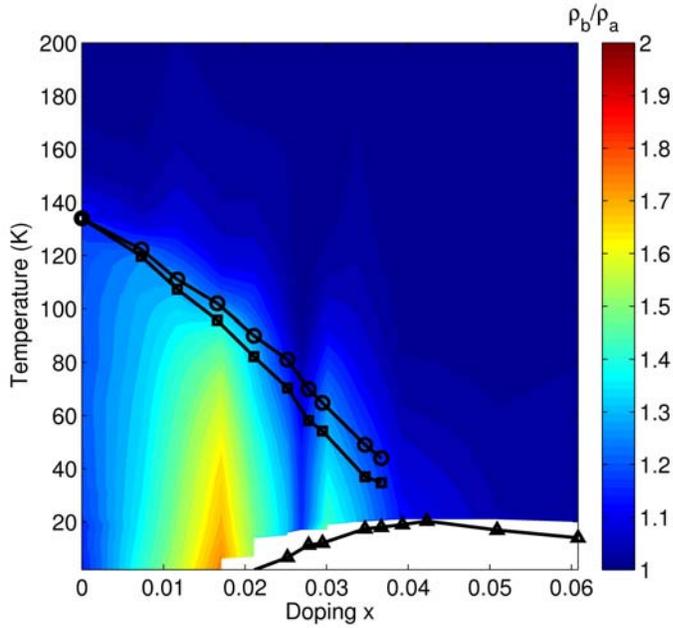

**Figure 13** Temperature and composition dependence of the in-plane resistivity anisotropy $\rho_b/\rho_a$ of $Ba(Fe_{1-x}Ni_x)_2As_2$ Structural, magnetic and superconducting critical temperatures, determined in the absence of uniaxial stress, are shown as circles, squares, and triangles, respectively. From Kuo *et al* (2011).

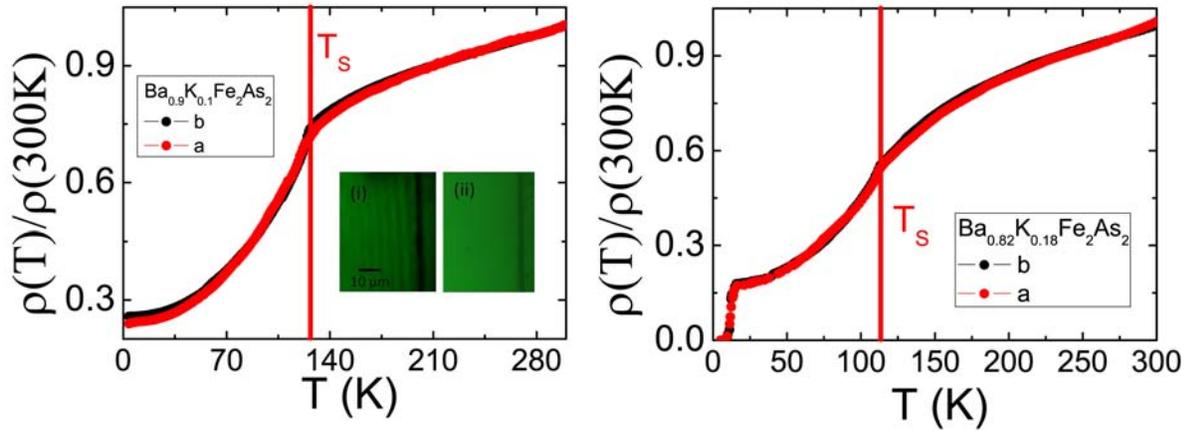

**Figure 14** Temperature dependence of the in-plane resistivity $\rho_a$ and $\rho_b$ for strained crystals of the hole-doped system $Ba_{1-x}K_xFe_2As_2$ for (a) $x=0.1$ and (b) $x = 0.18$. Insets to panel (a) show polarized light images which reveal the presence (absence) of twin domains for the unstressed (stressed) conditions. In stark contrast to the cases of $Ba(Fe_{1-x}Co_x)_2As_2$ and $Ba(Fe_{1-x}Ni_x)_2As_2$, no appreciable in-plane resistivity anisotropy is observed for $Ba_{1-x}K_xFe_2As_2$. Taken with permission from Ying *et al* 2010.

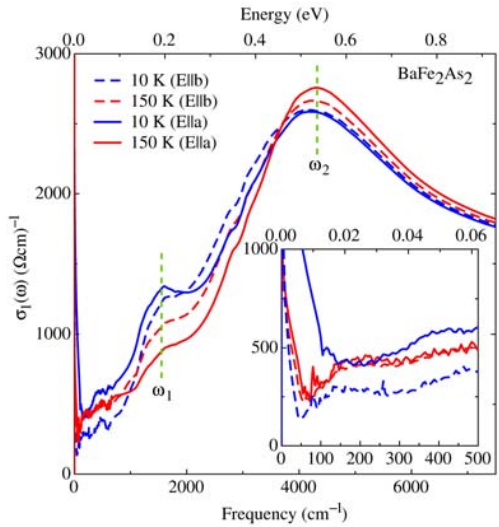

**Figure 15** Real part $\sigma_1(\omega)$ of the optical conductivity of $BaFe_2As_2$ at 10 and 150 K in the MIR-NIR range for both polarization directions. The vertical dashed lines mark the frequencies $\omega_1$ and $\omega_2$ (see text). Inset: Infrared part of $\sigma_1(\omega)$ at 10 and 150 K for both axes. From Dusza *et al* (2011).

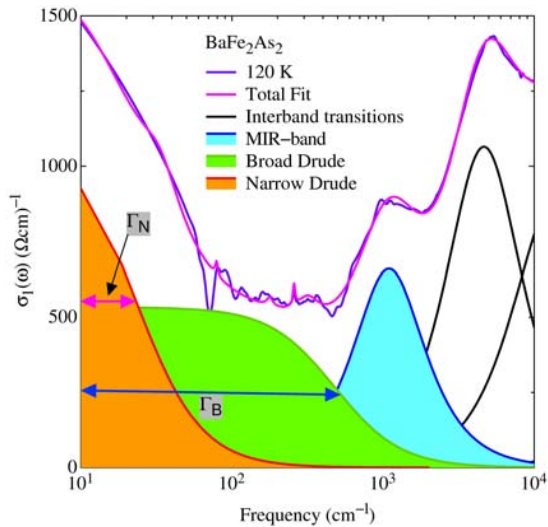

**Figure 16** Phenomenological modeling of the optical conductivity within the Drude-Lorentz approach: the narrow and broad Drude terms, the mid-infrared h.o. as well as the first and low-frequency tail of the second h.o. shaping $\sigma_1(\omega)$ above 2000 cm$^{-1}$ and reproducing the high frequency interband transitions. A third h.o. (not shown) finally allows reproducing the high frequency tail of the strong absorption in $\sigma_1(\omega)$ centered at about 4300 cm$^{-1}$. The colored shaded areas emphasize the spectral weights of both Drude terms (proportional to the squared plasma frequency) and MIR-band (proportional to the oscillator strength) [Dressel-Gruner-2002]. The total fit reproduces in great detail the measured spectrum (here as an example the parent compound $BaFe_2As_2$ at 120 K). This fitting approach applies to all temperatures and doping levels [Lucarelli *et al* 2010] and can be easily adapted to both polarization directions in detwinned materials [Dusza *et al* 2011].

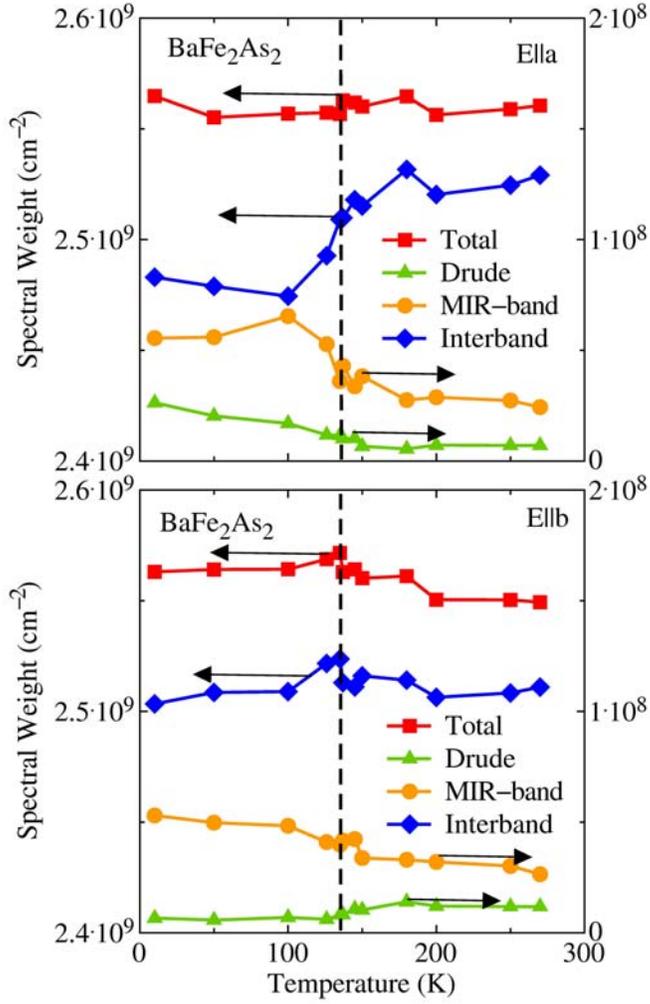

**Figure 17** Temperature dependence of the spectral weight (SW) redistribution along the *a*- and *b*-axis for *x*=0, extracted from the Drude-Lorentz fit (Fig. 16) of the optical response [Dusza *et al* 2011, Lucarelli *et al* 2010]. Both panels display SW encountered in both Drude terms, in the MIR-band and in the h.o.'s reproducing the high frequency interband transitions, as well as the resulting total SW. The vertical dashed line in both panels marks the phase transition at $T_N$.

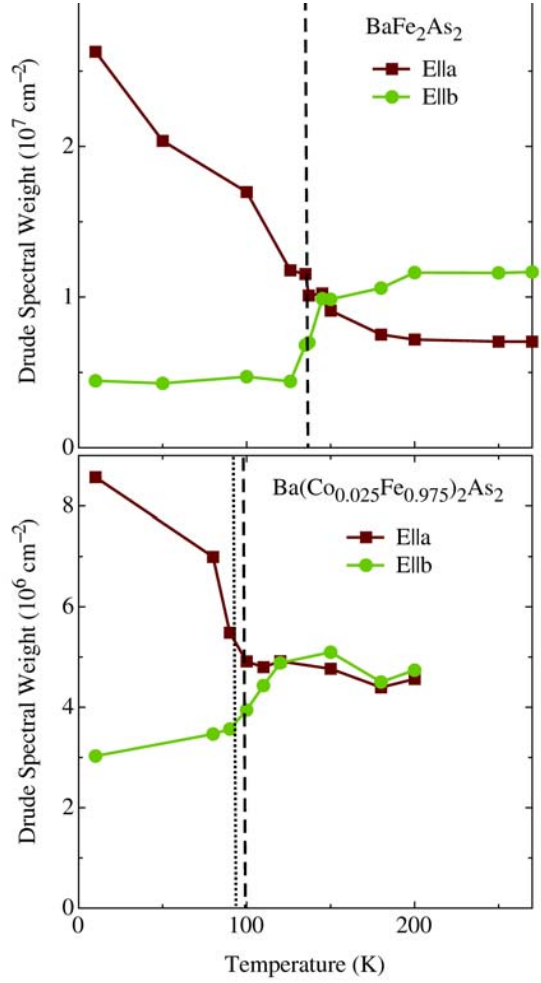

**Figure 18** Comparison of the total Drude spectral weight (sum of narrow and broad terms) of Ba(Fe$_{1-x}$Co$_x$)$_2$As$_2$ for $E \parallel a$ and $E \parallel b$ as a function of temperature for (a) $x = 0$ and (b) $x = 0.025$. Vertical lines mark $T_N$ and $T_s$.

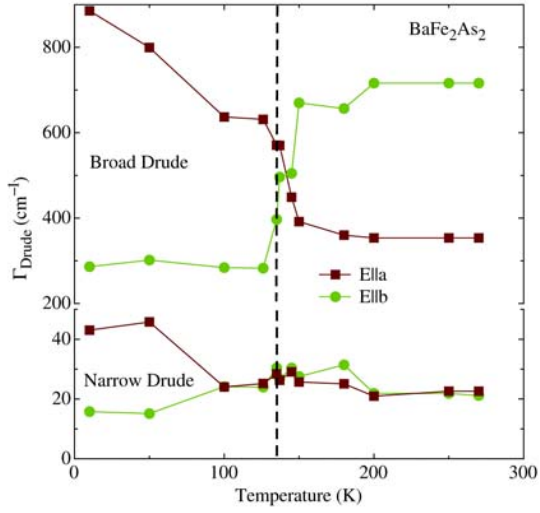

**Figure 19** Scattering rates $\Gamma_N$ and $\Gamma_B$ (Fig. 16) along the *a*- and *b*-axis for x=0, extracted from the Drude-Lorentz fit of the optical response [Dusza *et al* 2011, Lucarelli *et al* 2010]. The vertical dashed line marks the phase transition at $T_N$.

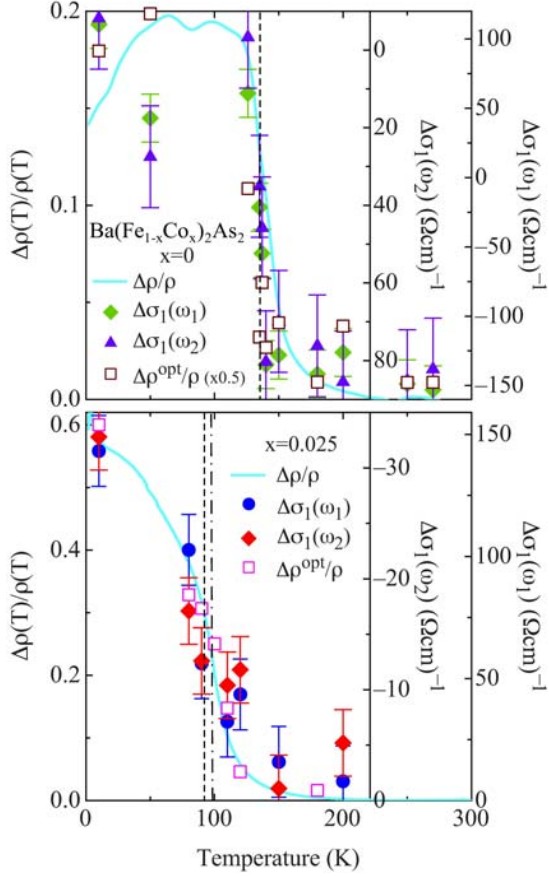

**Figure 20** Temperature dependence of the dichroism $\Delta\sigma_1(\omega)$ for *x*=0 and 0.025 at $\omega_1$ and $\omega_2$ (Fig. 15) compared to $\Delta\rho/\rho$ from the *dc* transport data [Chu *et al* 2010b], as well as from the Drude terms in $\sigma_1(\omega)$ ($\Delta\rho^{opt}/\rho$). The vertical dashed and dashed-dotted lines mark the magnetic and structural phase transitions at $T_N$ and $T_s$, respectively for unstressed conditions [Dusza *et al* 2011].

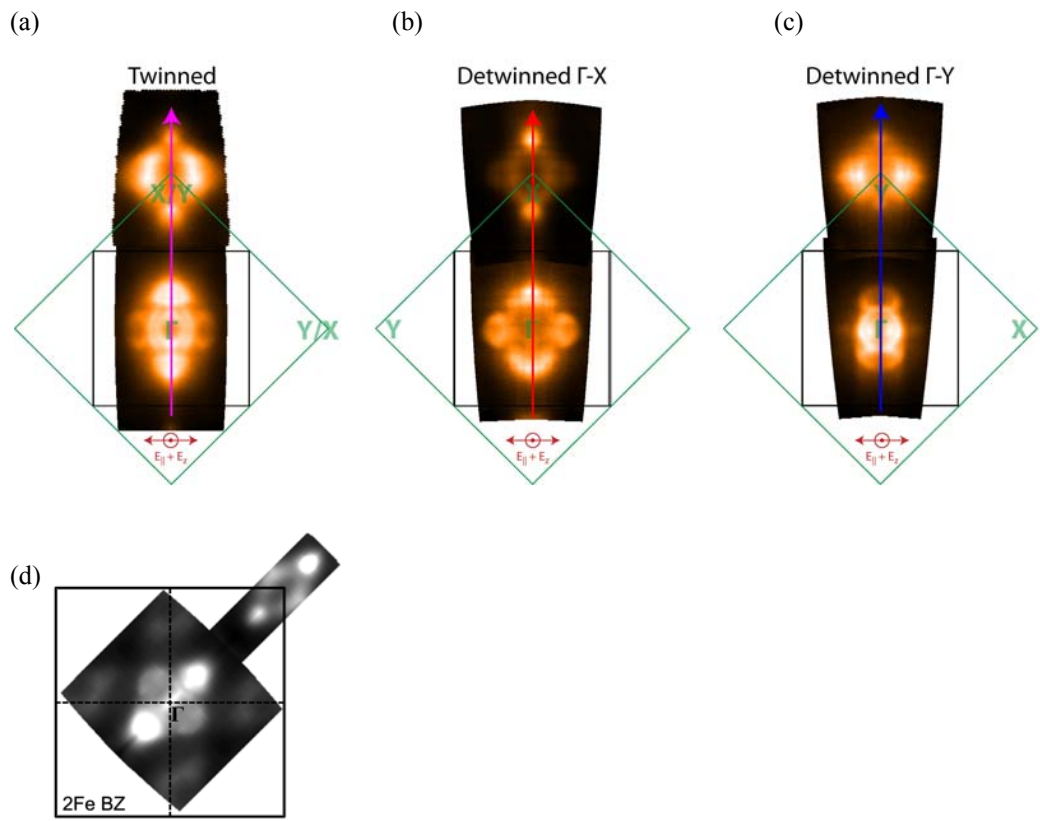

**Figure 21** The Fermi surface of (a) twinned and (b,c) untwinned crystals of $BaFe_2As_2$ in the SDW state. For twinned samples measurements reveal a superposition of the electronic structure observed for the detwinned crystals. Light polarization as indicated in each panel. From Yi *et al* (2011). (d) Similar results have also been obtained by Kim *et al* (2011) for detwinned single crystals, shown here along the $\Gamma X$ direction (the axes are rotated by 45 degrees with respect to panel b). Reprinted with permission from Y K Kim *et al* (2011), copyright (2011) by the American Physical Society.

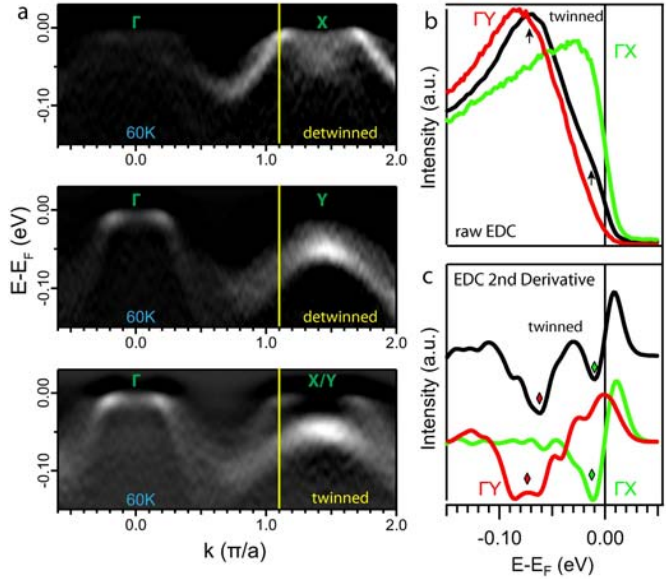

**Figure 22** (a) Second derivatives of spectral images taken along the Γ-X and Γ-Y high symmetry directions (upper panels) of detwinned crystals of Ba(Fe$_{0.975}$Co$_{0.025}$)$_2$As$_2$ taken at 60 K with 62 eV photons reveal a clear splitting of hole-like bands close to the X and Y points, identified from polarization dependence as being principally $d_{yz}$ and $d_{xz}$ character respectively. Similar measurements of twinned crystals (lower panel) reveal a superposition of the two upper panels. (b) EDCs at the momentum marked by yellow line of the raw spectral images corresponding to the spectral images shown in (a). (c) Second derivatives of the EDCs shown in panel (b), revealing that the splitting of the two bands for stressed (detwinned) and unstressed (twinned) crystals is the same within the resolution of the measurement. The curve for the twinned crystal is offset for clarity. From Yi *et al* (2011).

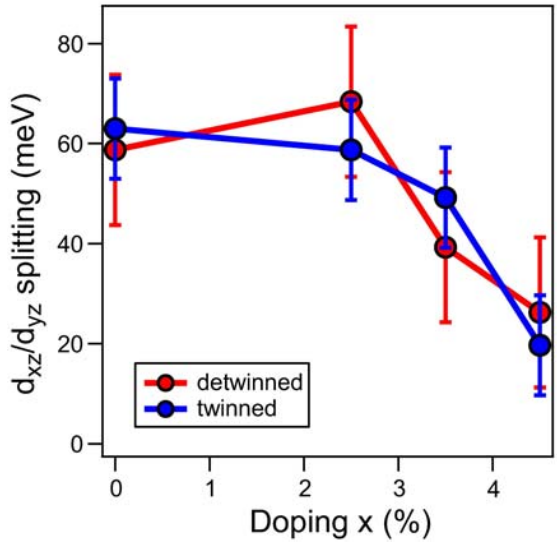

**Figure 23** Doping dependence of the splitting of the $d_{yz}$ and $d_{xz}$ bands for twinned and detwinned single crystals of Ba(Fe$_{1-x}$Co$_x$)$_2$As$_2$ measured at momentum k = 0.9 π/a taken at 10 K, with 47.5 eV photons ($k_z$ = 0). The energy splitting is monotonically suppressed with increasing Co substitution, and is the same for stressed (detwinned) and unstressed (twinned) crystals within the resolution of the measurement. From Yi *et al* (2011).

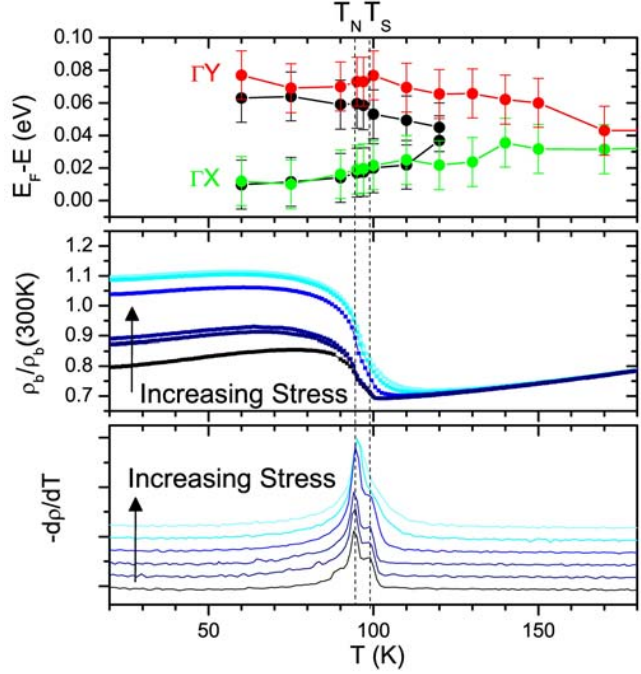

**Figure 24** Temperature dependence of (a) the energy of the $d_{yz}$ and $d_{xz}$ bands for twinned (black data points) and detwinned (red and green data points) crystals, (b) the resistivity, and (c) the temperature derivative of the resistivity, for single crystals of Ba(Fe$_{0.975}$Co$_{0.025}$)$_2$As$_2$. The data indicate a remarkable sensitivity to uniaxial stress of the electronic structure and the associated resistivity anisotropy for temperatures above $T_s$. (Figure modified from data shown in Yi *et al* (2011) and Chu *et al* (2010b).)